\documentclass[12pt]{article}

\usepackage{amssymb,amsmath}

\usepackage{graphicx}
\usepackage{color}
\usepackage[colorlinks=true
,urlcolor=blue
,citecolor=blue
,linkcolor=blue
,pagecolor=blue
,linktocpage=true
,pdfproducer=medialab
]{hyperref}
\usepackage[a4paper,width=15.5cm]{geometry}
\usepackage{subfigure}

\makeatletter \renewcommand{\@dotsep}{10000} \makeatother

\def\tb{\tilde b}

\def\tst{\tilde t}

\def\gev{\,{\rm GeV}}
\def\tev{\,{\rm TeV}}

\def\brhgm{$(Br/Br_{SM})_{h \rightarrow \gamma \gamma} $}
\def\brhz{$(Br/Br_{SM})_{h \rightarrow ZZ} $}

\newcommand{\beq}{\begin{equation}}
\newcommand{\eeq}{\end{equation}}
\newcommand{\bea}{\begin{eqnarray}}
\newcommand{\eea}{\end{eqnarray}}

\begin{document}

\begin{center}

 {\Large\bf

Higgs Boson Production and Decay: Effects from Light Third Generation and Vectorlike Matter

 } \vspace{1cm}

{\large   M. Adeel Ajaib\footnote{ E-mail: adeel@udel.edu}, Ilia Gogoladze\footnote{E-mail: ilia@bartol.udel.edu\\
\hspace*{0.5cm} On  leave of absence from: Andronikashvili Institute
of Physics, 0177 Tbilisi, Georgia.} and  Qaisar Shafi\footnote{ E-mail:
shafi@bartol.udel.edu} } \vspace{.9cm}

{\baselineskip 20pt \it
Bartol Research Institute, Department of Physics and Astronomy, \\
University of Delaware, Newark, DE 19716, USA  } \vspace{.5cm}

\vspace{1.5cm}
 {\bf Abstract}
\end{center}

 We study the implications of light third generation sparticles on the production cross section and decay widths of a light CP-even Higgs boson.  For simplicity, we consider scenarios in which only one of the sfermions from the third generation is light. For each case, we attempt to explain the apparently large enhancement in the Higgs production and decay in the diphoton channel with small deviations in the $ZZ$ channel. In the MSSM framework we find that  only a light stau can explain these observations while keeping the lightest CP-even Higgs boson mass in the interval 123 GeV $\lesssim m_h \lesssim$ 127 GeV. For the light stop scenario, the observations related to the diphoton and $ZZ$ channel
can be accommodated but, in order to satisfy the Higgs mass bound,  one needs to go beyond the MSSM. In particular, we invoke vector like particles with masses around a TeV. These new particles preserve gauge coupling unification and provide additional contributions to the Higgs mass. With these new contributions a 126 GeV Higgs mass is easily achieved. We also find that with only a light  sbottom quark, the above mentioned excess is hard to accommodate.

\newpage

\renewcommand{\thefootnote}{\arabic{footnote}}
\setcounter{footnote}{0}



\section{Introduction  \label{intro}}

The ATLAS and CMS collaborations at the Large Hadron Collider (LHC)  have independently reported the discovery \cite{ATLAS:jul4,CMS:jul4} of a particle with production and decay modes that seem more or less consistent with the Standard Model (SM) Higgs boson with a mass of around 126 GeV.  Complementary evidence is also provided by the updated combination of the
 Higgs searches  performed by the CDF and D0 collaborations at the Tevatron \cite{Tevatron:2012zz}.
 Understanding the properties of this boson is crucial and may direct us to the favored scenario for physics beyond the SM.
 A Higgs mass of around 126 GeV already applies  stringent  constraints on the various   supersymmetric (SUSY) models \cite{Baer:2011ab,Draper:2011aa}.
 In the decoupling limit ($m_A \gg m_Z$), the lightest  CP-even Higgs $h$ in the Minimal Supersymmetric SM (MSSM) has SM-like properties. Here $m_A$ denotes the mass of the CP-odd Higgs boson in the MSSM. The MSSM can accommodate the value $m_h \sim 126 \rm \ GeV$, but this  requires either very large, $O (\mathrm{few}-10)$ TeV,  stop quark mass, or a large trilinear soft supersymmetry  breaking (SSB) A-term, with a stop quark mass of around a TeV \cite{Baer:2011ab,Draper:2011aa}.
 It was shown in   ref. \cite{Gogoladze:2011aa} that, assuming gravity mediated SUSY breaking \cite{Chamseddine:1982jx}, a SM-like Higgs boson with mass 
$\sim 126  {\rm \ GeV}$ is nicely accommodated in SUSY  grand unified theory (GUT) models with $t$-$b$-$\tau$ Yukawa coupling unification at $M_{\rm GUT}$ \cite{big-422}.

In addition to the Higgs discovery the ATLAS and CMS experiments have both observed an excess in Higgs production and decay in the diphoton  channel which is a factor $1.4-2$ times larger than the SM expectations. For the final state consisting of a pair of $Z$ bosons, the ATLAS experiment sees an excess, whereas CMS observes a deficit. However, both are currently consistent with the presence of a SM Higgs boson \cite{CMS:ZZ-4l,ATLAS:ZZ-4l}. The observed signal for these channels is quantified by the ratio of the product
of production cross sections times branching ratio to the final state $XX$ compared to the theoretical expectation for the SM. Thus,
\begin{eqnarray}
R_{XX} \equiv \frac{\sigma(h) \times Br(h\rightarrow XX)}{(\sigma(h) \times Br(h\rightarrow XX))_{SM}}.
\label{eq:ratio}
\end{eqnarray}
The current values of this ratio for the $\gamma \gamma$ and $ZZ$ channels are follows:
\begin{equation}
\begin{tabular}{lll}
ATLAS: & $R_{\gamma \gamma}=1.90 \pm 0.5\, ,$  & $R_{ZZ}=1.3 \pm 0.6 \, ,$  \\
CMS: & $R_{\gamma \gamma}=1.56 \pm 0.43 \, , $  & $R_{ZZ}=0.7 \pm 0.5 \, ,$ \\
ATLAS$\oplus$CMS: & $R_{\gamma \gamma}=1.71 \pm 0.33 \, , $  & $R_{ZZ}=0.95
\pm 0.4 \, .$
\end{tabular}
\label{eq:masterequation}
\end{equation}
Note that the combination of the  ATLAS and CMS  results is taken from   ref. \cite{Arbey:2012dq}.
The present deviations observed in these channels, if they persist, will provide strong evidence for physics beyond the SM.

In this paper we attempt to explain the  observations  presented in Eq. (\ref{eq:masterequation})  with a MSSM spectrum consisting of light third generation squarks or sleptons.  Compared to previous studies \cite{ Djouadi:1998az, Desai:2012qy, Carena:2011aa}, we require that the observations related to both $R_{\gamma \gamma}$ and
$R_{ZZ}$  are satisfied simultaneously with a Higgs mass of around 126 GeV.
As previously mentioned, a Higgs mass $\sim$  126 GeV typically needs very heavy stop quarks.
It was shown in ref. \cite{Moroi:1991mg,Gogoladze:2009bd} that introducing vector like particles at the TeV scale with suitable couplings to the MSSM Higgs can provide a significant contribution to the lightest CP-even Higgs mass.
We will see that the presence of vector like particles can allow the stop quarks to be as light as the current experimental bound.
 The presence of a relatively light stop quark then enables us to explain the observations presented in Eq.(\ref{eq:masterequation}). We find  that the light stop scenario is capable  not only of accommodating the combined ATLAS and CMS observations but also the current seperate observations from ATLAS and CMS for the diphoton and $ZZ$ channels  which, of course, needs  future confirmation.

The layout of this  paper is as follows. In Section \ref{higgs-prod} we briefly review the production and decay of the Higgs via the process $gg \rightarrow h \rightarrow \gamma \gamma$, and discuss conditions that can enhance or suppress the $\sigma \times Br$. Section \ref{constraintsSection} outlines the scanning procedure and the phenomenological constraints that we apply.  In section \ref{sec:only-light-stop} we consider, in the decoupling limit, a scenario in which stop is the next to lightest SUSY particle and discuss the implications of this on the Higgs production cross section and branching ratios. In section \ref{sec:light-stop} we discuss the light stop case when the mass of the pseudo-scalar Higgs $m_A$ is relatively small. In section \ref{sec:light-sbottom} we consider the light sbottom case for relatively low values of $m_A$. We discuss light stau scenario, similar to the stop case, in sections \ref{sec:only-light-stau} and \ref{sec:light-stau}. Our conclusions are presented in Section \ref{conclusions}.

\section{$gg \rightarrow h \rightarrow \gamma \gamma$ Process \label{higgs-prod}}

\subsection{$gg \rightarrow h$}
The gluon fusion process is the main production channel of the Higgs at the LHC. In the SM, the leading order (LO) process involves  a top quark loop which has the largest Yukawa coupling with the Higgs. The cross-section for this process is known to the next-to-next-to-leading order (NNLO) \cite{Anastasiou:2002yz} which can enhance the LO result by 80-100$\%$. Any new particle which strongly couples with the Higgs can significantly enhance this cross-section. In the MSSM the stop plays such a role and therefore this process can probe the stop sector with the exception of scenarios when the contribution from sbottom becomes important.  The decay width for this process is given by (see \cite{Djouadi:2005gj,gunion90} and references therein)
\begin{eqnarray}
\Gamma(h \to gg) &=& \frac{G_F \alpha_s^2 m_h^3}{36 \sqrt{2}\pi^3}
       \left| N_{c}\, Q_{t}^{2}\,  g_{htt}\, A_{\frac12}^h(\tau_t) +  {\cal A}_{\text{\tiny{SUSY}}  }^{gg}
     \right|^2 ,
\end{eqnarray}
where $g_{htt}$ is the coupling of $h$ to the top quark and $\tau_i = m_h^2/(4m_i^2)$. The form factors are given by
\begin{eqnarray}
A_{\frac12}^h(\tau) &=& \frac2{\tau^2} [ \tau + (\tau - 1)f(\tau)]\ , \\
A_0^h(\tau) &=& -\frac{1}{\tau^2}\left[ \tau - f(\tau) \right] \ , \\
A_1^h(\tau) &=&  -\frac{1}{\tau^2}[2\tau^2+3\tau+3(2\tau-1)f(\tau) ] \ , \\
\label{ftau}
f(\tau) &=& \left\{ \begin{array}{lc}
       \displaystyle     \arcsin^2 \sqrt{\tau} & \quad \tau \le 1 \\
  \displaystyle -\frac14 \left[ \log
    \frac{1+\sqrt{1-\tau^{-1}}}{1-\sqrt{1-\tau^{-1}}} - i \pi \right]^2
      & \quad \tau > 1
         \end{array} \right. 
\end{eqnarray}
The supersymmetric contribution ${\cal A}_{  \text{\tiny{SUSY}}  }^{gg}$ is given by
\begin{eqnarray}
 {\cal A}_{  \text{\tiny{SUSY}}  }^{gg} &=&  \sum_{i} N_{c}\, Q_{\tilde{q}_i}^{2}\, g_{h\tilde{q}_i\tilde{q}_i}\, \frac{m_Z^2}{m_{\tilde{q}_i}^2}\,
                 A_{0}^h(\tau_{\tilde{q}_i})\ .
\end{eqnarray}
The couplings $g_{h\tilde{q}_i\tilde{q}_i}$ of the CP-even Higgs boson to the squark mass eigenstates, normalized to $2(\sqrt{2}G_F)^{1/2}$, are given by \cite{Djouadi:2005gj,gunion90, Boonekamp:2005up}
\small{\begin{eqnarray}
  g_{h \tilde t_1 \tilde t_1} & = &  \left( \frac{1}{2} c^2_{ \theta_{\tilde
t}}
    - \frac{2}{3} s_w^2 c_{ 2 \theta_{\tilde t} }\right)\! M^2_Z \sin
(\beta + \alpha)
    - m^2_t\, \frac{\cos \alpha}{\sin \beta} - \frac{1}{2} s_{ 2 \theta_{\tilde t} }\, m_t
    \left( A_t \frac{\cos \alpha}{\sin \beta} + \mu \frac{\sin \alpha}{\sin
\beta} \right),\hskip .7cm \nonumber  \\
  g_{h \tilde t_2 \tilde t_2} & = &  \left( \frac{1}{2} s^2_{ \theta_{\tilde
t}}
    + \frac{2}{3} s_w^2 c_{ 2 \theta_{\tilde t} } \right)\! M^2_Z \sin
(\beta + \alpha)
    - m^2_t\, \frac{\cos \alpha}{\sin \beta} + \frac{1}{2} s_{ 2 \theta_{\tilde t} }\, m_t
    \left( A_t \frac{\cos \alpha}{\sin \beta} + \mu \frac{\sin \alpha}{\sin
\beta} \right), \nonumber \\
  g_{h \tilde b_1 \tilde b_1} & = & \left( \frac{1}{2} c^2_{ \theta_{\tilde b}}
    - \frac{1}{3} s_w^2 c_{ 2 \theta_{\tilde b} } \right)\! M^2_Z \sin
(\beta + \alpha)
    - m^2_b\, \frac{\sin \alpha}{\cos \beta} - \frac{1}{2} s_{ 2 \theta_{\tilde b} }\, m_b
    \left(  A_b \frac{\sin \alpha}{\cos \beta} + \mu \frac{\cos \alpha}{\cos
\beta} \right),  \nonumber \\
  g_{h \tilde b_2 \tilde b_2} & = & \left( \frac{1}{2} s^2_{ \theta_{\tilde b}}
    + \frac{1}{3} s_w^2 c_{ 2 \theta_{\tilde b} } \right)\! M^2_Z \sin
(\beta + \alpha)
    - m^2_b\, \frac{\sin \alpha}{\cos \beta} + \frac{1}{2} s_{ 2 \theta_{\tilde b} }\, m_b
    \left(  A_b \frac{\sin \alpha}{\cos \beta} + \mu \frac{\cos \alpha}{\cos
\beta} \right),\nonumber
\end{eqnarray}}
\normalsize
where $s_w \equiv \sin^2 \theta_W$, $c_{\theta} \equiv \cos \theta$ and $\theta_{\tilde{q}}$ is the mixing angle between the flavor basis and mass eigenbasis. The couplings for the stau have expressions similar to that of the sbottom with the relevant electric charge for the stau in the first parenthesis. The cross section for the $gg \rightarrow h$ process is directly proportional to the decay width $\Gamma(gg \rightarrow h)$. The stop and sbottom loop contribution goes like $ 1/m_{\tilde{q}}^2$ and can significantly enhance the cross section for light squarks. Moreover the cross section can also increase from an enhancement in the couplings  $g_{h\tilde{q}_i\tilde{q}_i}$. The latter enhancement can arise due to light stops, large values of the mixing parameter  $A_{t}$ and also large $\mu \tan\beta$. We shall discuss the enhancement and suppression of this cross section in more detail in the following sections where we present our results.

\subsection{$h \rightarrow \gamma \gamma$}
The Higgs boson can decay to a pair of gauge bosons, leptons, or quarks. The dominant decay channel for a 126 GeV Higgs is a  pair of b quarks ($b \bar b$)  at  tree level,  but is not very useful due to the large QCD background. One of the most promising decay channels is the $\gamma \gamma$ final state which, at leading order, proceeds through a loop containing charged particles, namely the charged Higgs, sfermions and charginos. The dominant contribution to $h \rightarrow \gamma \gamma$ decay comes from the W boson loop and the decay width is given by (see \cite{Djouadi:2005gj,gunion90} and references therein)
\begin{eqnarray}
\Gamma(h\rightarrow\gamma\gamma)&=&\frac{G_{F}\alpha^{2}m_{h}^{3}}{128\sqrt{2}\pi}
\left| N_{c}\, Q_{t}^{2}\, g_{htt}\, A_{1/2}^{h}(\tau_{t})+
g_{hWW}\, A_{1}^{h}(\tau_{W}) + {\cal A}_{  \text{\tiny{SUSY}}  }^{\gamma\gamma}\right|^2,
\end{eqnarray}
where $g_{hWW}$ is the coupling of $h$ to the $W$ boson. The supersymmetric contribution $ {\cal A}_{  \text{\tiny{SUSY}}  }^{\gamma\gamma} $ is given by
\begin{eqnarray}
{\cal A}_{  \text{\tiny{SUSY}}  }^{\gamma\gamma} &=&
 g_{hH^{+}H^{-}}\, \frac{m_{W}^{2}}{m^{2}_{H^{\pm}}} \, A_{0}^{h}(\tau_{H^{\pm}}) +
\sum_f N_c Q_f^2\, g_{h\tilde{f}\tilde{f}}\, \frac{m_Z^2}{m^2_{\tilde{f}}}\, A_0^h(\tau_{\tilde{f}})
     + \nonumber \\
&& \quad \quad \quad \sum_i g_{h\chi_i^+\chi_i^-}\, \frac{m_W}{m_{\chi_i}}\, A_{\frac12}^h(\tau_{\chi_i}),
\end{eqnarray}
where $g_{hXX}$ is the coupling of $h$ to the particle $X$ ($= H^{\pm}, \tilde{f}, \chi^{\pm}_i$). The stop and sbottom loop factors have similar contributions as the gluon fusion case. In this case however the stau can also contribute to enhance the decay width without changing the gluon fusion cross section. The chargino contribution to the decay width is known to be less than 10\% for $m_{\chi_{i}^{\pm}} \gtrsim 100 \rm \ GeV$. The charged Higgs contribution is even smaller since its coupling to the CP-even Higgs is not proportional to its mass and also due to the loop suppression $m_{W}^{2}/m^{2}_{H^{\pm}}$. For a light stop the Higgs production and decay can be significantly enhanced. For a light sbottom the enhancement in the gluon fusion production can be large but an overall enhancement in $gg \rightarrow h \rightarrow \gamma \gamma$ is difficult to achieve as we shall see in our analysis.


\section{Phenomenological Constraints and Scanning Procedure\label{constraintsSection}}
We employ the FeynHiggs 2.9.0~\cite{feynhiggs} package to perform random
scans over the MSSM fundamental parameter space.  The range of the parameters we choose in each case are given in subsequent sections.
In our analysis the first and second generations are decoupled since their masses are assumed to be around 5 TeV. The gaugino mass parameters $M_1, \ M_2$ and $ M_3$ are also taken to be 5 TeV.
 We set the top quark mass  $m_t = 173.3\, {\rm GeV}$  \cite{:1900yx}. The version of FeyHiggs we employ also tests for color and charge breaking (CCB), and therefore points where color breaking minima is detected are rejected.

In performing the random scan a uniform and logarithmic distribution of random points is first generated in the selected parameter space.
The function RNORMX \cite{Leva} is then employed
to generate a gaussian distribution around each point in the parameter space.  The points with $0.8< R_{XX}< 3$ are scanned more rigorously using this function.

 We successively apply the following experimental constraints on the data that
we acquire from FeynHiggs: $m_{\tst_1}> 130\gev$ \cite{Aad:2011ib,He:2011tp},   $\ m_{\tb_1} >100 \gev$ \cite{arXiv:1103.4344, AdeelAjaib:2011ec}, $m_{\tilde{\tau_1}}> 105\gev$, \cite{Amsler:2008zzb}.
The  lower bound on sfermion masses are consistent with nearly degenerate neutralino-sfermion scenarios, which are very helpful in obtaining the correct relic abundances \cite{Jungman:1995df}. We do not apply constraints from B-physics in our analysis since our aim is to highlight the effects of a  light third generation on the Higgs production and decay to the $\gamma \gamma $ and $ZZ$ final states. In each scenario we choose our parameters to make one of the sparticles from the third generation light with all others decoupled so that there effects on B-physics are negligible. In principle other sparticles can also be  light and hence give contributions to B-physics. However, such an analysis would involve additional parameters in each case and therefore require a much more extensive analysis.

\begin{figure}[]
\newpage
\vspace{-1cm}

\centering
\subfigure[]{\label{fig:ost-a}{\includegraphics[scale=0.4]{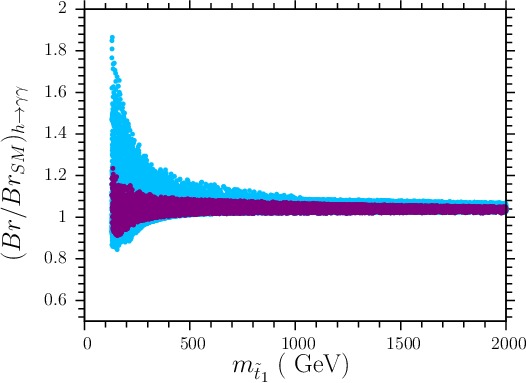}}}\hfill
\subfigure[]{\label{fig:ost-b}{\includegraphics[scale=0.4]{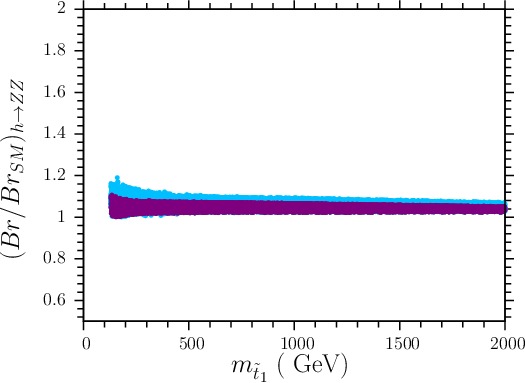}}}\\
\subfigure[]{\label{fig:ost-c}{\includegraphics[scale=0.4]{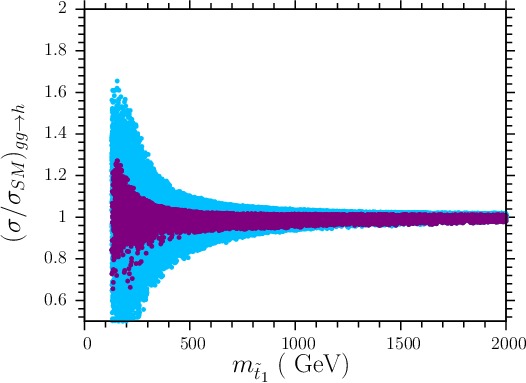}}}\hfill
\subfigure[]{\label{fig:ost-d}{\includegraphics[scale=0.4]{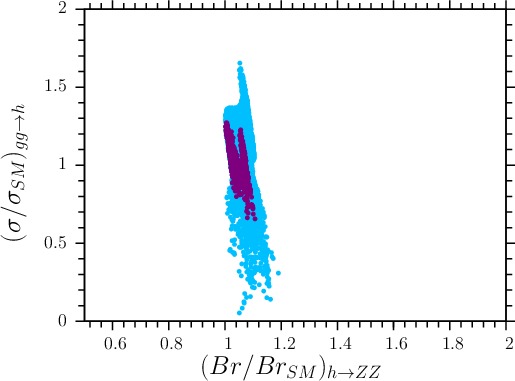}}}\\
\subfigure[]{\label{fig:ost-e}{\includegraphics[scale=0.41]{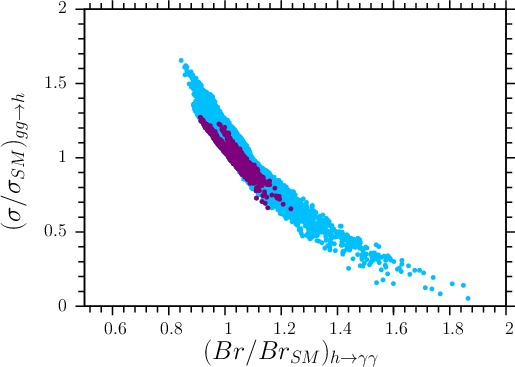}}}\hfill
\subfigure[]{\label{fig:ost-f}{\includegraphics[scale=0.41]{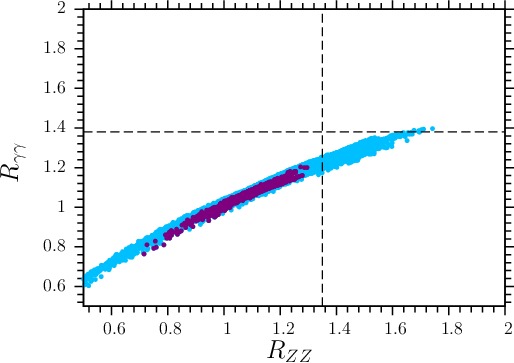}}}
\caption{Plots in the $Br/Br_{\rm SM}$ vs. $m_{\tilde{t}_1}$ plane for \subref{fig:ost-a} $h \rightarrow \gamma \gamma$ and  \subref{fig:ost-b} $h \rightarrow ZZ$ channels. Panel \subref{fig:ost-c} shows the ratio of the cross section $\sigma / \sigma_{\rm SM}$ vs. $m_{\tilde{t}_1}$ for the gluon fusion process. The ratio of the cross section and branching ratio for the $h \rightarrow \gamma \gamma$ and $h \rightarrow ZZ$ vs. $gg \rightarrow h$ channel are plotted in panels \subref{fig:ost-d} and \subref{fig:ost-e}. Panel  \subref{fig:ost-f} shows the plot of the product $R_{ \gamma \gamma}$ vs. $R_{ ZZ}$, where $R$ is defined in Eq. (\ref{eq:ratio}). The purple points satisfy the Higgs mass window given in Eq. (\ref{eq:mh-range}). The vertical dashed line in panel \subref{fig:lst-f} shows the upper bound on $R_{ ZZ}$ and lower bound on $R_{ \gamma \gamma}$ from the combined analysis given in Eq. (\ref{eq:masterequation}). All points satisfy the constraints described in section \ref{constraintsSection}.}
\label{fig:only-light-stop1}
\end{figure}



\begin{figure}[]
\begin{center}
\includegraphics[scale=0.4]{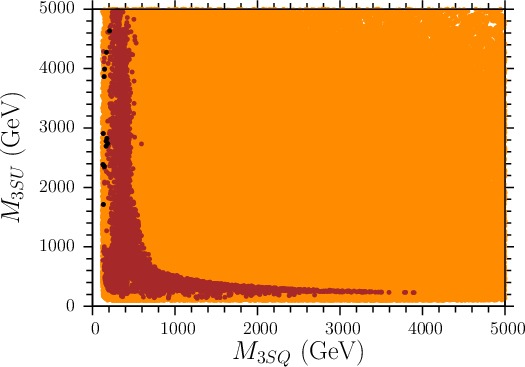}
\includegraphics[scale=0.4]{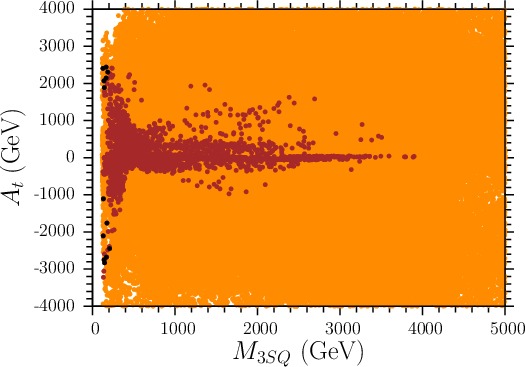}\vspace*{3mm}
\end{center}
\caption{Plots in the $M_{3SU}$ vs. $M_{3SQ}$ and $A_t$ vs. $M_{3SQ}$ planes. Orange points satisfy the constraints described in section \ref{constraintsSection}. The brown points form a subset of the orange points and satisfy the current limits on $R_{gg}$ and $R_{ZZ}$ from the CMS experiment given in Eq. (\ref{eq:masterequation}). The black points form a subset of the brown points that satisfy the Higgs mass range given in Eq. (\ref{eq:mh-range}).}
\label{fig:only-light-stop2}
\end{figure}


\section{ Light Stop in the Decoupling limit \label{sec:only-light-stop}}

We first consider a scenario with only a pair of light scalar top quarks effectively  contributing  to new physics via Higgs production and decay processes. We assume the decoupling limit ($m_A \gg m_Z$) in which the lightest Higgs is SM-like and the other Higgs bosons are nearly degenerate ($m_A \simeq m_H \simeq m_{H}^{\pm}$). For this case we scan the following range of the parameter space,
\begin{eqnarray}
   & 100\gev < M_{3SU}, M_{3SQ} < 5000 \gev, \nonumber \\
   & {-4000\ \gev}< A_t < 4000\  {\rm GeV},\nonumber \\
& 3 < \tan\beta  < 60,
\label{eq:param1}
\end{eqnarray}
where, $M_{3SQ}, M_{3SU}$ are the mass parameters of the third generation left handed squark doublet and right handed top squark, respectively.  The parameter $A_{t}$ is the coefficient of the trilinear soft term associated with the top quark Yukawa coupling. All other A terms are set equal to zero. ${\tan\beta}$ is the
ratio of the VEVs of the two MSSM Higgs doublets.  We assume the neutralino to be the lightest supersymmetric particle (LSP) which is nearly degenerate  with the light stop quark. This assumption relaxes the stop mass bound compared with other colored sparticles \cite{He:2011tp}. 

 In Figure \ref{fig:ost-a} and \ref{fig:ost-b},  we show our results  in the $Br/Br_{\rm SM}$ vs. $m_{\tilde{t}_1}$ planes for the  $h \rightarrow \gamma \gamma$ and $h \rightarrow ZZ$ decay channels. The cross section ratio $\sigma / \sigma_{\rm SM}$ vs. $m_{\tilde{t}_1}$ for the gluon fusion process is shown in Figure \ref{fig:ost-c}.
The ratio of the $gg \rightarrow h$ cross section is plotted versus the branching ratio of the $h \rightarrow \gamma \gamma$ and $h \rightarrow ZZ$ channels in Figures \ref{fig:ost-d} and \ref{fig:ost-e}. The ratio $R$ is plotted in Figure  \ref{fig:ost-f} for the $h \rightarrow \gamma \gamma$ vs. $h \rightarrow ZZ$ channel and is given by Eq. (\ref{eq:ratio}).
All the points displayed in Figure \ref{fig:only-light-stop1} satisfy the experimental constraints described in section \ref{constraintsSection}. The points shown in purple in Figure \ref{fig:only-light-stop1} satisfy the following Higgs mass window
\begin{eqnarray}
123 \ \mathrm{GeV} \le m_h \le 127 \ \mathrm{GeV}.
\label{eq:mh-range}
\end{eqnarray}

The first notable feature in these figures is the large enhancement of the diphoton production and gluon fusion process in Figure \ref{fig:ost-a} and \ref{fig:ost-c}.
It has been noted before \cite{Djouadi:1998az} that for a light stop and small $A_t$, the gluon fusion rate can be enhanced by up to 60\% due to constructive interference of the stop and top loops in the gluon fusion cross section. The diphoton decay, however, is suppressed by up to 20\% due to destructive interference of the W boson and top/stop loops. Together, this leads to an overall enhancement in the product $R_{\gamma \gamma}$. For large values of the parameter $A_t$  the gluon fusion cross section is suppressed due to destructive interference between the top and stop loops. This cancellation leads to enhancement in the diphoton channel which is now dominated by the W boson loop as  seen in Figure \ref{fig:ost-a}. The reduction in the gluon fusion rate however is much stronger, so that the overall enhancement  in $R_{ \gamma \gamma}$ is not large.  On the other hand, the purple points show that the large enhancement in the diphoton production and gluon fusion process through the light stop contribution is drastically reduced once the Higgs mass bound from Eq. (\ref{eq:mh-range}) is applied to the data.
 Figure \ref{fig:ost-b} shows that the enhancement in the $ZZ$ production is not large compared to the diphoton case. This is because in the decoupling limit, the coupling of the CP-even Higgs with the gauge boson is proportional to $g \sin(\beta - \alpha) \sim 0$.

From Figure \ref{fig:ost-d} we can see that the gluon fusion cross section does not vary significantly with change in the branching ratio to a pair of $Z$ bosons. Figure \ref{fig:ost-e}, however, shows that the gluon fusion cross section has an inverse relationship with  the diphoton branching ratio. This trend shows that the overall enhancement in $R_{\gamma \gamma}$ does not become large over the whole region of the parameter space for this scenario. The reason for this inverse trend is that the enhancement in the gluon fusion rate is from the constructive interference of the top and stop loops, which is accompanied by the cancellation of these with the W boson loop. The enhancement in the diphoton rate, which is due to destructive interference between the top and stop loops, is accompanied by a reduction in the gluon fusion rate. We can notice that the reduction in the gluon fusion rate is much stronger for relatively larger values of the diphoton decay rate.

The ATLAS and CMS experiments have seen an enhancement in the $\gamma \gamma$ final state which is $1.4-2$ times  the SM value. The enhancement seen by ATLAS is accompanied by an enhancement in the $ZZ$ final states, whereas this is not the case for CMS, as can be seen from the current limits given in Eq.  (\ref{eq:masterequation}). Clearly, more data is required to settle this. If the enhancement in the $\gamma \gamma$ channel is accompanied by an enhancement in the $ZZ$ channel the light stop scenario is disfavored. Figure \ref{fig:ost-f} shows that an enhancement in $R_{ZZ}$ is  accompanied by a similar but less stronger enhancement in  $R_{\gamma \gamma}$  for values greater than 1. The dashed lines show the upper bound on $R_{ZZ}$ and the lower bound on $R_{\gamma \gamma}$ for the combined analysis given in Eq. (\ref{eq:masterequation}).  If this  bound from  the ATLAS and CMS collaborations is confirmed in the near future,  it will rule out the light stop scenario.

In Figure \ref{fig:only-light-stop2} our results are shown in the $M_{3SU}$ vs. $M_{3SQ}$ and $A_t$ vs. $M_{3SQ}$ planes. The orange points show the data that is consistent with the bounds discussed in section \ref{constraintsSection}. 
 The brown points also form a subset of the orange points satisfying the current limits on the diphoton and $ZZ$ channels from the  CMS given in Eq.  (\ref{eq:masterequation}).
The points shown in black form a subset of the brown points satisfying the limit on the Higgs mass given in Eq. (\ref{eq:mh-range}). As seen from the figures and also described above, the CMS observations seem to be  in favor of the light stop scenario. We can observe a large region of the parameter space consistent with the CMS bound (brown points), whereas there are no points satisfying the ATLAS bound. This is because the central values of ATLAS indicate an enhancement in both the $\gamma \gamma$ and $ZZ$ channels, which is not favored in this scenario as seen in Figure \ref{fig:ost-d}. We may also note that the Higgs mass constraints is  satisfied by the very few points shown in black. This shows that requiring the Higgs mass to be $\sim 126 \gev$ appears to disfavor this scenario. However, as we will discuss in the next section, the contributions of vector-like matter to the Higgs mass can ameliorate this situation.
%


\begin{figure}[]
\centering
\subfigure[]{\label{fig:lst-a}{\includegraphics[scale=0.4]{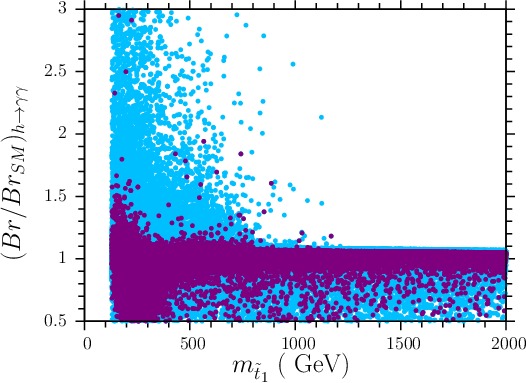}}}\hfill
\subfigure[]{\label{fig:lst-b}{\includegraphics[scale=0.4]{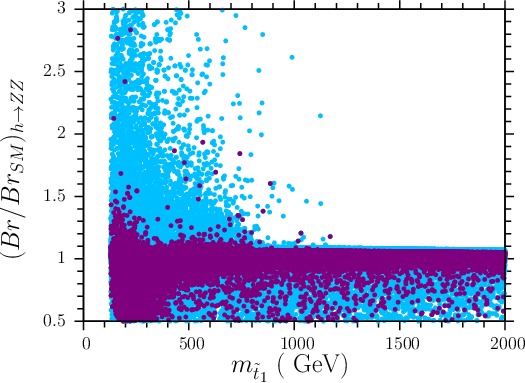}}}\\
\subfigure[]{\label{fig:lst-c}{\includegraphics[scale=0.4]{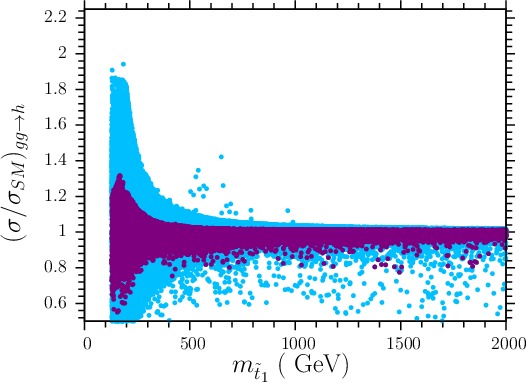}}}\hfill
\subfigure[]{\label{fig:lst-d}{\includegraphics[scale=0.41]{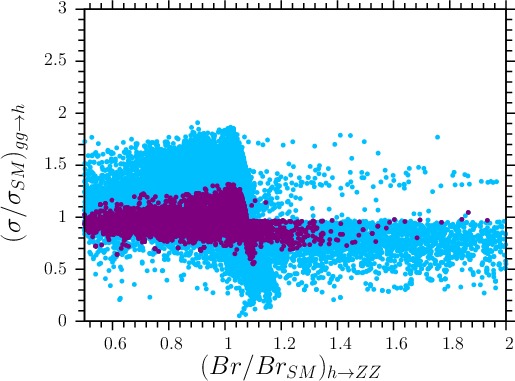}}}\\
\subfigure[]{\label{fig:lst-e}{\includegraphics[scale=0.41]{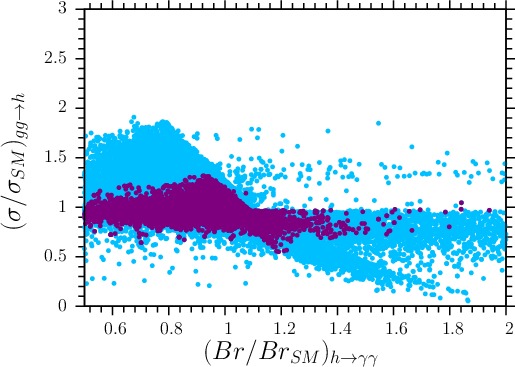}}}\hfill
\subfigure[]{\label{fig:lst-f}{\includegraphics[scale=0.41]{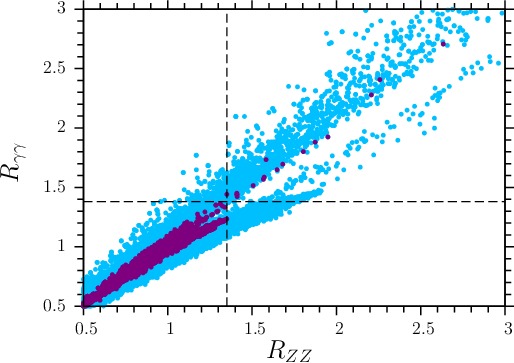}}}
\caption{Plots in the $(Br/Br_{\rm SM})_{h \rightarrow \gamma \gamma}$ vs. $m_{\tilde{t}_1}$, $(Br/Br_{\rm SM})_{h \rightarrow ZZ}$ vs. $m_{\tilde{t}_1}$  and $(\sigma / \sigma_{\rm SM})_{gg \rightarrow h}$ vs. $m_{\tilde{t}_1}$ planes.
  The ratio of the cross sections and branching ratio for the $gg \rightarrow h$ vs. $h \rightarrow \gamma \gamma$ and $h \rightarrow ZZ$  channels are plotted in panels \subref{fig:lst-d} and  \subref{fig:lst-e}. Panel  \subref{fig:lst-f} shows the plot in the $R_{ \gamma \gamma}$ vs. $R_{ ZZ}$ planes.  The color coding and definition of the dashed lines is given in Figure \ref{fig:only-light-stop1}.}
\label{fig:light-stop1}
\end{figure}



\begin{figure}[]
\centering
\includegraphics[scale=0.4]{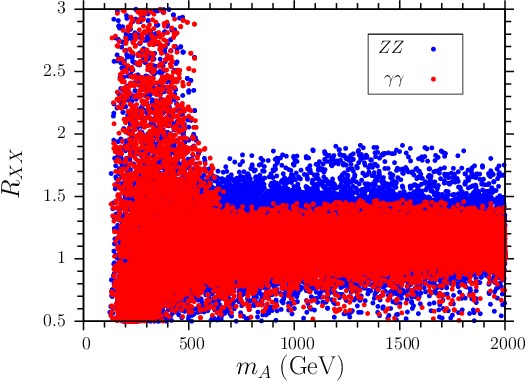}
\includegraphics[scale=0.4]{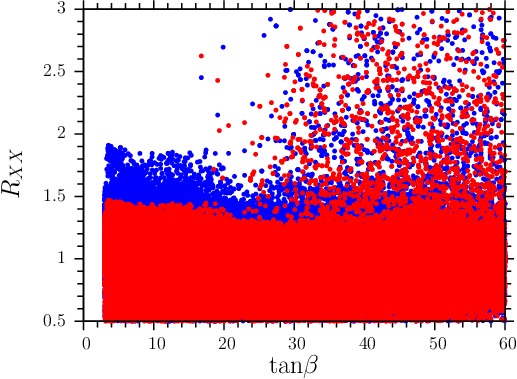}\vspace*{3mm}
\caption{$R_{XX}$ vs. $m_A$ and $R_{XX}$ vs. $\tan\beta$ planes. The red  points correspond to $R_{\gamma \gamma}$ and the blue points correspond to $R_{ZZ}$.}
\label{fig:light-stop3}
\end{figure}



\begin{figure}[]
\centering
\subfigure[]{\label{fig:lst-2a}{\includegraphics[scale=0.4]{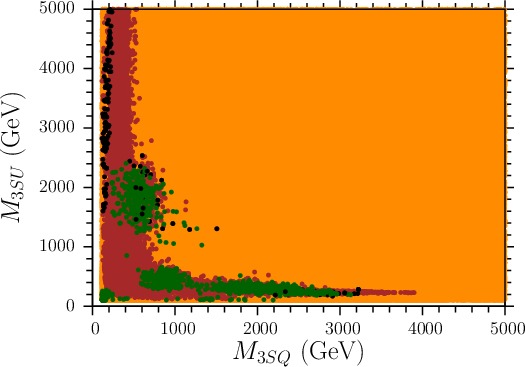}}}\hfill
\subfigure[]{\label{fig:lst-2b}{\includegraphics[scale=0.4]{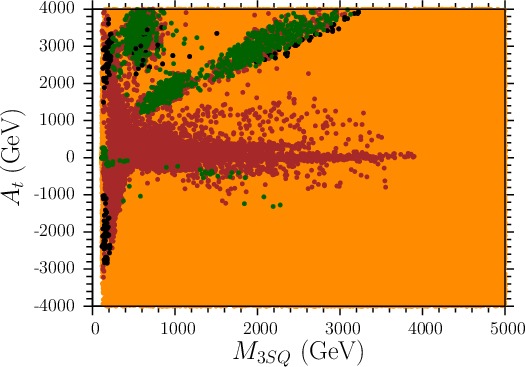}}}\\
\subfigure[]{\label{fig:lst-2c}{\includegraphics[scale=0.4]{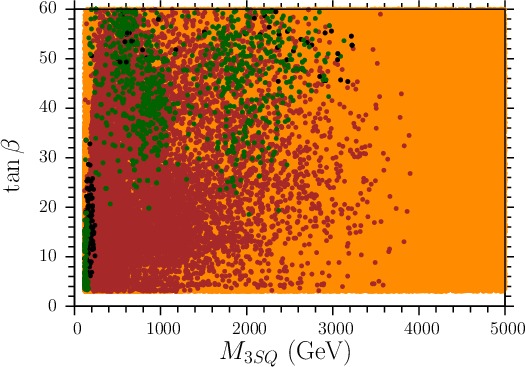}}}\hfill
\subfigure[]{\label{fig:lst-2d}{\includegraphics[scale=0.4]{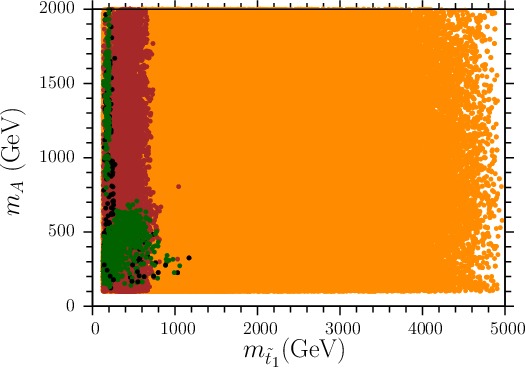}}}
\caption{Plots in the $M_{3SU}$ vs. $M_{3SQ}$, $A_t$ vs. $M_{3SQ}$, $\tan\beta$ vs. $M_{3SQ}$ and $m_A$ vs. $m_{\tilde{t}_1}$ planes. The green and brown points form a subset of the orange points and satisfy the current limits on $R_{gg}$ and $R_{ZZ}$ from the ATLAS and CMS experiments given in Eq. (\ref{eq:masterequation}). The black points form a subset of the green and brown points and satisfy the Higgs mass bounds given in Eq. (\ref{eq:mh-range}).}
\label{fig:light-stop2}
\end{figure}




\begin{figure}[]
\begin{center}
\includegraphics[scale=0.3]{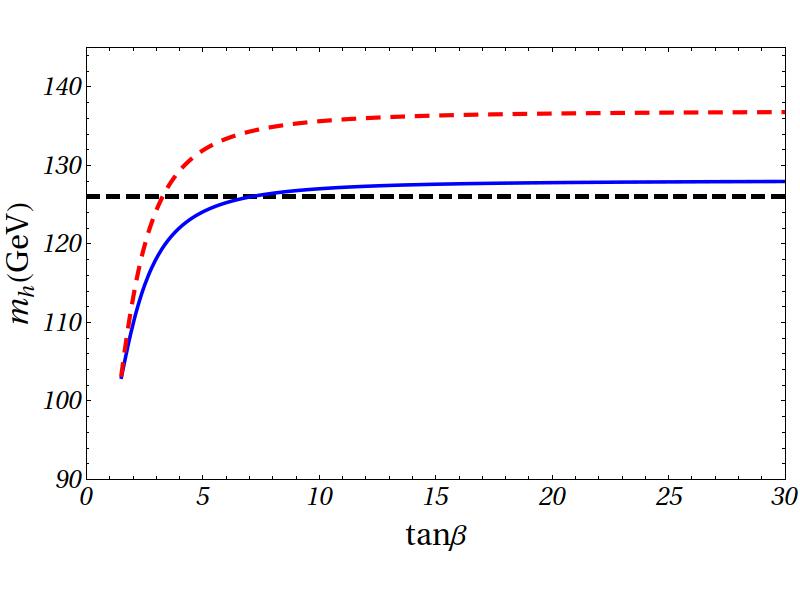}
\end{center}
\caption{$m_h$ vs. $\tan\beta$  plane illustrating the contributions of vector-like multilplets to the Higgs mass. The blue curve  corresponds to $M_S=2 \rm TeV$ and $X_t=6$, and the red dashed line corresponds to $(M_S,M_V,X_{k10},X_t)= (200 \rm {\ GeV}, 2 \rm {\ TeV}, 3, 6)$ and $\kappa_{10}=1$. The black dashed line shows $m_h= 126 \rm \ GeV$. }
\label{fig:vector-like}
\end{figure}



\section{Light Stop and low $m_A$ region \label{sec:light-stop}}

In this section we vary $\mu$ and $m_A$ in order to explore the light stop scenario in the region with low values of the CP-odd Higgs mass $m_A$. The decay width of the Higgs boson into a pair of b quarks can be modified due to low/moderate values of $m_A$, and this can have important effects on the other decay channels as well. The diphoton branching ratio is given as
\begin{eqnarray}
Br(h \rightarrow \gamma \gamma) \approx \frac{\Gamma(h \rightarrow \gamma \gamma)}{\Gamma(h \rightarrow b \overline{b})}.
\end{eqnarray}
For low/moderate values of $m_A$ and large/moderate $\tan\beta$, the $b \overline{b}$ and $\tau \overline{\tau}$ channels can be suppressed and this, in turn, can enhance the other decay channels. Similarly, an enhancement in the $h \rightarrow b \overline{b}$ channels leads to a suppression of the other decay channels.

For this case we scan the following range of the parameter space,
\begin{eqnarray}
   & 100\gev < M_{3SU}, M_{3SQ} < 5000 \gev, \nonumber \\
   & {-4000\gev}< A_t < 4000  \gev,\nonumber \\
 & 100\gev< m_A < 2000 \gev, \nonumber \\
   & 100\gev <  \mu < 1000 \gev, \nonumber \\
   & 3 < \tan\beta  < 60.  \nonumber
\label{eq:param12}
\end{eqnarray}

The first and second generation masses are assumed to be  5 TeV. All other A-terms are set to zero. Our results for this case are shown in Figures \ref{fig:light-stop1} and \ref{fig:light-stop2}. In Fig \ref{fig:light-stop1} we plot the same variables as in Figure \ref{fig:only-light-stop1}. Comparing the figures for the two cases we can notice a much broader region showing enhancement in the $\gamma \gamma$ and $ZZ$ final states. Figures \ref{fig:lst-a}, \ref{fig:lst-b} show that this enhancement in the $\gamma \gamma$ and $ZZ$ final states can now accommodate much larger stop masses ($m_{\tilde{t}_1} \lesssim 1 \tev$) compared to the decoupling limit. In other words, heavier stops can now accommodate the enhancement and also satisfy the Higgs mass range from Eq.(\ref{eq:mh-range}), as seen by the broader coverage of the purple points in this Figure. The enhancement in the cross section in Figure \ref{fig:lst-c} shows a similar trend as in the previous case and corresponds to small values of $A_t$, resulting from the destructive interference of the stop and top loops.  Figures \ref{fig:lst-d} and \ref{fig:lst-e}  show that for smaller cross section, there are points with larger branching ratio for the two decay channels. In the decoupling case we saw an inverse trend between the $Br$ and cross section, which is not present in this case due to additional enhancement for low values of $m_A$.  Figure \ref{fig:lst-f} again plots the measurable quantities $R_{\gamma \gamma}$ vs. $R_{ ZZ}$. We can see  a large number of data points above the dashed lines and therefore a much broader region is able to satisfy the current bounds on these products.

We also observe from Figure \ref{fig:lst-f} that an enhancement in $R_{\gamma \gamma}$ can be explained by a light stop.  The product $R_{ \gamma \gamma}$ is significantly enhanced  for a light stop quark mass, which makes it difficult to get the correct Higgs mass in this scenario. It was noticed  in refs. \cite{Moroi:1991mg,Gogoladze:2009bd} that in the presence of a vector like particles around the TeV region and with suitably large  couplings to the Higgs field, one can have sizable corrections to the light CP-even Higgs mass.
As an example, particles which are in the  $10+\overline{10}$ dimensional representation of the SU(5) symmetry group
were introduced.  In the superpotential, the coupling $\kappa_{10}10\, 10\, 5_{H}$ contains the interaction $\kappa_{10} Q_{10}\, U_{10}\, H_{u}$. Here $Q_{10}$ and $U_{10}$ stand for vector like particles which have the same MSSM quantum numbers as the left and right handed up type quarks. $H_{u}$ is the MSSM up type Higgs field and $\kappa_{10}$ is a dimensionless coupling. In this case  the
CP-even Higgs boson gets the following additional contribution to its mass \cite{Moroi:1991mg,Gogoladze:2009bd}:
\begin{eqnarray}
\left[ m_{h}^{2}\right] _{10} &=&-M_{Z}^{2}\cos ^{2}2\beta \left(
\frac{3}{8\pi ^{2}}\kappa_{10} ^{2}t_{V}\right)
+\frac{3}{4\pi ^{2}}\kappa_{10} ^{4}v^{2}\sin ^{2}\beta \left[ t_{V}+\frac{1%
}{2}X_{\kappa_{10} }\right].  \label{eq1}
\end{eqnarray}
Here  $X_{\kappa_{10} }$ and $t_V$  are given as follows
\begin{equation}
X_{\kappa_{10} }=\frac{4\widetilde{A}_{\kappa_{10} }^{2}\left(
3M_{S}^{2}+2M_{V}^{2}\right) -\widetilde{A}_{\kappa_{10}
}^{4}-8M_{S}^{2}M_{V}^{2}-10M_{S}^{4}}{6\left(
M_{S}^{2}+M_{V}^{2}\right) ^{2}}, \label{f1}
\end{equation}%
and
\begin{equation}
t_{V}=\log \left( \frac{%
M_{S}^{2}+M_{V}^{2}}{M_{V}^{2}}\right), \label{mm7}
\end{equation}
 where $\widetilde{A}_{\kappa_{10}}=A_{\kappa_{10}}-\mu \cot \beta $,
$A_{\kappa_{10}}$ is the $Q_{10}-U_{10}$ trilinear soft mixing
parameter and $\mu$ is the MSSM Higgs bilinear mixing term.
$M_S\simeq \sqrt{m_{\tilde{Q}_{3}}\,m_{\tilde{U}_{3}^c}}$, where
$m_{\tilde{Q}_{3}}$ and $m_{\tilde{U}_{3}^c}$ are the stop left and
stop right soft SUSY breaking masses at low scale. $M_V$ is the mass term for the vector like particles.

The leading  1- and 2- loop
contributions to the lightest CP-even Higgs boson mass in the MSSM is given by
\cite{at, Carena:1995wu}%
\begin{eqnarray}
\left[ m_{h}^{2}\right] _{MSSM} &=&M_{Z}^{2}\cos ^{2}2\beta \left( 1-\frac{3%
}{8\pi ^{2}}\frac{m_{t}^{2}}{v^{2}}t\right)  \nonumber \\
&+&\frac{3}{4\pi ^{2}}\frac{m_{t}^{4}}{v^{2}}\left[ t+\frac{1}{2}X_{t}+\frac{%
1}{\left( 4\pi \right) ^{2}}\left(
\frac{3}{2}\frac{m_{t}^{2}}{v^{2}}-32\pi \alpha _{s}\right) \left(
X_{t}t+t^{2}\right) \right],  \label{eq2}
\end{eqnarray}
where
\\
\begin{eqnarray}
t =\log \left(
\frac{M_{S}^{2}}{M_{t}^{2}}\right),~
X_{t} &=&\frac{2\widetilde{A}_{t}^{2}}{M_{S}^{2}}\left( 1-\frac{\widetilde{A}%
_{t}^{2}}{12M_{S}^{2}}\right). \label{A1}
\end{eqnarray}%
Also $\widetilde{A}_{t}=A_{t}-\mu \cot \beta $, where
$A_{t}$ denotes the stop left and stop right soft
mixing parameter. The total CP-even Higgs mass is therefore given by
\begin{eqnarray}
m_h^2= \left[ m_{h}^{2}\right] _{MSSM} + \left[ m_{h}^{2}\right]
_{10}. \label{max_10}
\end{eqnarray}
In Figure \ref{fig:vector-like} we plot the mass $m_h$ vs. $\rm tan\beta$  for the MSSM and the MSSM + vector like particle  cases.
The blue curve corresponds to the  upper bound for the CP-even Higgs  mass if $M_S=2$ TeV and $A_t$ takes its  maximum possible values. It  hardly reaches the 126 GeV mass bound. On the other hand, in order to minimize the stop quark contribution to $m_h$ we could choose $M_S=200$ GeV and consider vector like particles with masses around  2 TeV. We choose $\kappa_{10}=1$ and $X_{\kappa_{10}}=3$. The
red dashed line shows that in this case the CP-even Higgs mass can be as large as 138 GeV.
This shows that in the presence of vector like particles we can have a stop quark as light as needed,  without worrying about the  value  of the lightest CP-even Higgs mass. Therefore, in the presence of  vector like particles the blue points in
Figure \ref{fig:lst-f} can accommodate the bounds from the ATLAS and CMS experiments.

 In Figure \ref{fig:light-stop3} our results are shown in the $R_{XX}$ vs. $m_A$ and $R_{XX}$ vs. $\rm tan\beta$ planes in order to emphasize the contribution from the MSSM CP-odd Higgs $A$. The red points show the product $R_{XX}$ for the $\gamma \gamma$ final state, whereas the points in blue show this for the $ZZ$ final state.  The additional enhancement observed in this case in Figures \ref{fig:lst-a} and \ref{fig:lst-b} corresponds to low values of $m_A \lesssim 600 \gev$ and $\tan\beta \gtrsim 30$. It has been discussed in earlier references \cite{Carena:1998gk, Carena:1999bh} that lower/moderate values of $m_A$ and $\tan\beta$ can suppress the $b \overline{b}$ and $\tau \tau$ channels and, as a result, the decays to $\gamma \gamma$ and $ZZ$ can be significantly enhanced. The sensitivity of $Br(h \rightarrow b \overline{b})$ to $m_A$ comes through the coupling $g_{hbb} \propto -\sin\alpha/\cos\beta$, where the mixing angle $\alpha$ is a function $m_A$. Moreover,  the radiative corrections to the Yukawa couplings of the b quarks and $\tau$ leptons (which are employed in FeynHiggs) can suppress these couplings significantly for large $ \mu \  \tan\beta$.

In Figure \ref{fig:light-stop2} we show plots of the fundamental parameters in the $M_{3SU}$ vs. $M_{3SQ}$, $A_t$ vs. $M_{3SQ}$, $\tan\beta$ vs. $M_{3SQ}$ and $m_A$ vs. $m_{\tilde{t}_1}$ planes. The orange, green and black points satisfy the same conditions as described in section \ref{sec:only-light-stop}. A much wider expanse of the parameter space now satisfies the current bounds from the ATLAS and CMS experiments given in Eq. (\ref{eq:masterequation}). The combination of the two experiments is also satisfied as can be seen from the overlap of the two regions. There are also more points shown in black satisfying the Higgs mass for stop mass $\lesssim 1 \tev$.  This case therefore provides a much richer parameter space that can accommodate the current bounds from experiments. Hence a spectrum consisting of a light stop with low $m_A$ and large $\tan\beta$ with all other particles decoupled can provide a possibility of explaining the current experimental bounds.


\begin{figure}[]
\centering
\subfigure[]{\label{fig:lsb-a}{\includegraphics[scale=0.4]{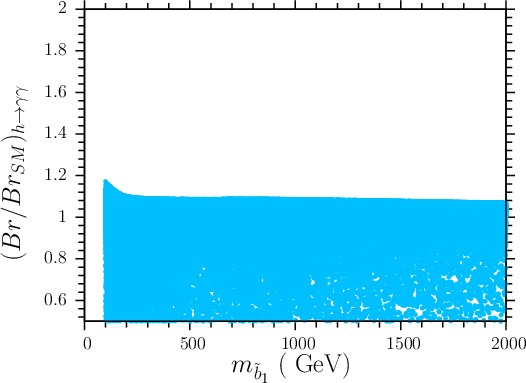}}}\hfill
\subfigure[]{\label{fig:lsb-b}{\includegraphics[scale=0.4]{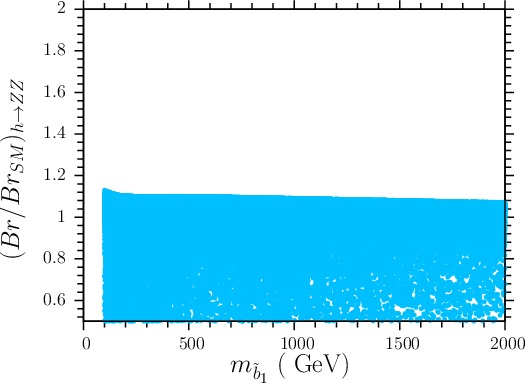}}}\\
\subfigure[]{\label{fig:lsb-c}{\includegraphics[scale=0.4]{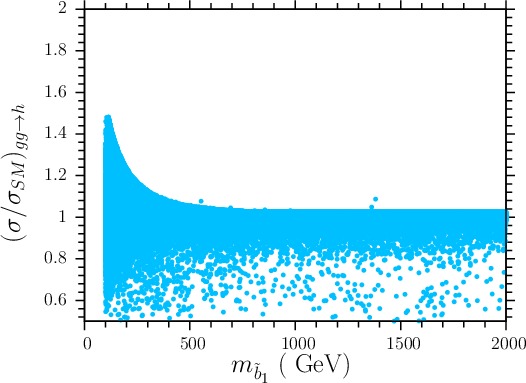}}}\hfill
\subfigure[]{\label{fig:lsb-d}{\includegraphics[scale=0.41]{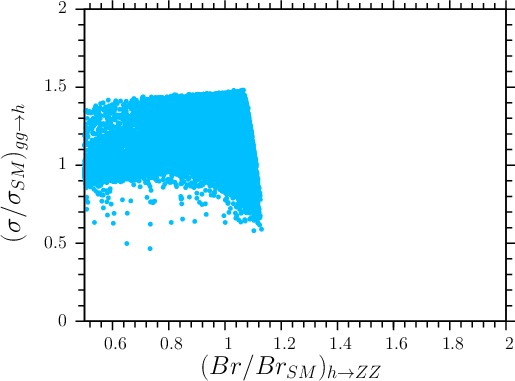}}}\\
\subfigure[]{\label{fig:lsb-e}{\includegraphics[scale=0.41]{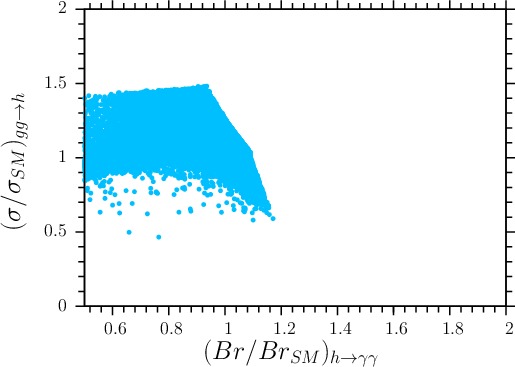}}}\hfill
\subfigure[]{\label{fig:lsb-f}{\includegraphics[scale=0.41]{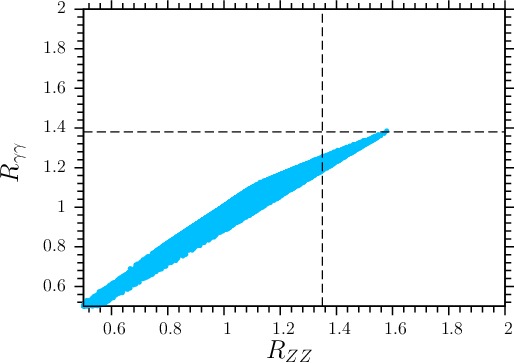}}}
\caption{$(Br/Br_{\rm SM})_{h \rightarrow \gamma \gamma}$ vs. $m_{\tilde{b}_1}$, $(Br/Br_{\rm SM})_{h \rightarrow ZZ}$ vs. $m_{\tilde{b}_1}$  and $(\sigma / \sigma_{\rm SM})_{gg \rightarrow h}$ vs. $m_{\tilde{t}_1}$ planes.
  The ratio of the cross section and branching ratio for  $gg \rightarrow h$ vs. $h \rightarrow \gamma \gamma$ and $h \rightarrow ZZ$  channels are plotted in panels \subref{fig:lsb-d} and  \subref{fig:lsb-e}. Panel  \subref{fig:lsb-f} shows the plot in the $R_{ \gamma \gamma}$ vs. $R_{ ZZ}$ planes. The definition of the dashed lines is given in Figure \ref{fig:only-light-stop1}.   All  points  satisfy the constraints described in section \ref{constraintsSection}. }
\label{fig:light-sbot1}
\end{figure}



\begin{figure}[]
\centering
\subfigure[]{\label{fig:lsb-2a}{\includegraphics[scale=0.4]{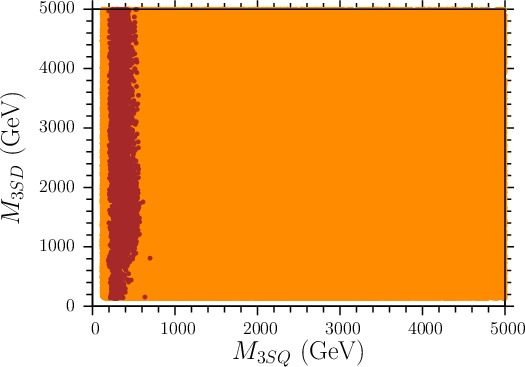}}}\hfill
\subfigure[]{\label{fig:lsb-2b}{\includegraphics[scale=0.4]{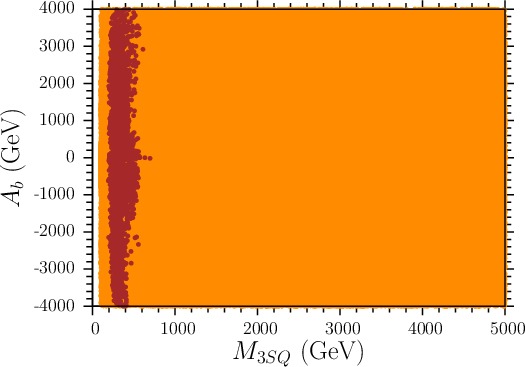}}}\\
\subfigure[]{\label{fig:lsb-2c}{\includegraphics[scale=0.4]{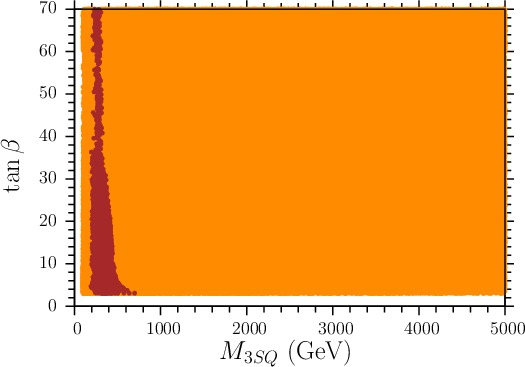}}}\hfill
\subfigure[]{\label{fig:lsb-2d}{\includegraphics[scale=0.4]{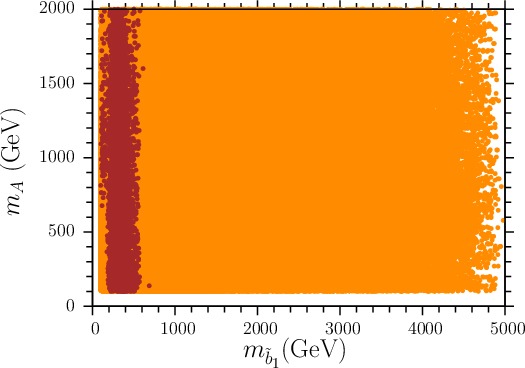}}}
\caption{Plots in $M_{SD3}$ vs. $M_{3SQ}$, $A_b$ vs. $M_{3SQ}$, $\tan\beta$ vs. $M_{3SQ}$ and $m_A$ vs. $m_{\tilde{b}_1}$ planes. Color coding is the same as in Figure \ref{fig:light-stop2}.}
\label{fig:light-sbot2}
\end{figure}



\section{Light Sbottom and low $m_A$ region\label{sec:light-sbottom}}

We next consider a scenario in which the spectrum consists of light bottom squarks and study its effects on the branching ratio and Higgs production cross section. For this case we scan the parameter space as follows:
\begin{eqnarray}
   & 100\gev < M_{3SD}, M_{3SQ} < 5000 \gev, \nonumber \\
   & {-4000 \gev}< A_{b} < 4000 \gev, \nonumber \\
 & 100\gev< m_A < 2000 \gev, \nonumber \\
   & 100\gev <  \mu < 1000 \gev, \nonumber \\
   & 3 < \tan\beta  < 60,
\label{eq:param13}
\end{eqnarray}
where $M_{3SD}$ is the mass parameter of the third generation right handed bottom squarks. As before, the first and second generation masses are assumed to be 5 TeV, and all the other A-terms are set to zero. In Fig \ref{fig:light-sbot1} we plot the same variables as in Figure \ref{fig:only-light-stop1} for the bottom squarks. The branching ratio for the $\gamma \gamma$ and $ZZ$ final states show a small enhancement for light sbottom mass in Figures \ref{fig:lsb-a} and \ref{fig:lsb-b}, whereas a strong enhancement in the cross section can be seen in Figure \ref{fig:lsb-c}. Note that the left handed squarks have the same mass due to $SU(2)$ gauge symmetry because of which the stop can also be light when the sbottom is light. Large difference in their masses can be achieved with large mixing in the stop sector. For the enhancement seen in Figure \ref{fig:lsb-c} the stop is also light so that its contribution also becomes important for this case.

 In Figures \ref{fig:lsb-d} and \ref{fig:lsb-e} we plot the gluon fusion vs. the branching ratio of the two channels. For increasing cross section the branching ratio is distributed over a continuous range of values that can either be small or large. The maximum value of the cross section ($\sim 1.5$) corresponds to  \brhgm   $\sim 0.9$ and \brhz $ \sim 1.1$ indicating maximum values of the products $R_{\gamma \gamma}$ $\sim 1.4$ and $R_{ZZ}$ $\sim 1.6$ as can be seen in Figure \ref{fig:lsb-f}. A linear correlation between $R_{\gamma \gamma}$ and $R_{ZZ}$ can be seen in Figure \ref{fig:lsb-f} as well. We can also notice from this figure that the diphoton channel is typically suppressed compared to the $ZZ$ channel, and this seems to be disfavored by current observations. From the dashed lines in this figure we can see that there is almost no parameter space that can satisfy the combined limit from the two experiments.

The fundamental parameters are plotted in Figure \ref{fig:light-sbot1} and show  the prospects for a light sbottom in the light of current limits from the ATLAS and CMS experiments. We notice a region of the parameter space that is consistent with the limits from CMS (brown points), whereas no agreement with the ATLAS experiment is seen. Based on the observations made from  Figure \ref{fig:lsb-f} we can see that the light sbottom scenario is also disfavored by the combination of the two experiments. The CMS  limits are satisfied for smaller values of the parameter $M_{3SQ}$, whereas values of the right handed sbottom mass $M_{3SD}$, the parameter $A_b$ and $\tan\beta$ cover the whole scanned range. Figure \ref{fig:lsb-2d} shows that the current limits from CMS prefer a sbottom with mass below $600 \gev$.



\begin{figure}[]
\centering
\subfigure[]{\label{fig:olta-a}{\includegraphics[scale=0.4]{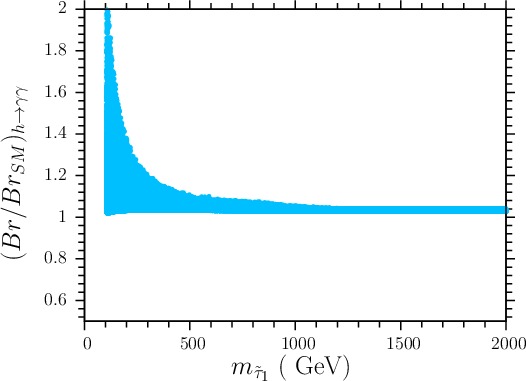}}}\hfill
\subfigure[]{\label{fig:olta-b}{\includegraphics[scale=0.4]{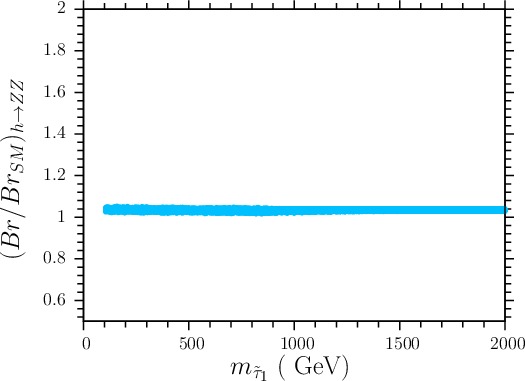}}}\\
\subfigure[]{\label{fig:olta-f}{\includegraphics[scale=0.41]{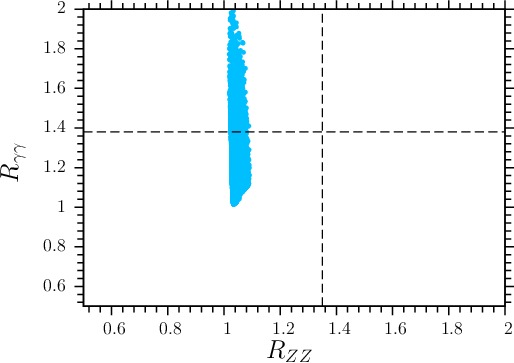}}}
\caption{$(Br/Br_{\rm SM})_{h \rightarrow \gamma \gamma}$ vs. $m_{\tilde{t}_1}$, $(Br/Br_{\rm SM})_{h \rightarrow ZZ}$ vs. $m_{\tilde{t}_1}$  and $(\sigma / \sigma_{\rm SM})_{gg \rightarrow h}$ vs. $m_{\tilde{t}_1}$ planes.
 Panel  \subref{fig:olta-f} shows the plot in the $R_{ \gamma \gamma}$ vs. $R_{ ZZ}$ planes.  The definition of the dashed lines is given in Figure \ref{fig:only-light-stop1}. All points satisfy the constraints described in section \ref{constraintsSection}.}
\label{fig:only-light-stau1}
\end{figure}



\begin{figure}[]
\begin{center}
\includegraphics[scale=0.4]{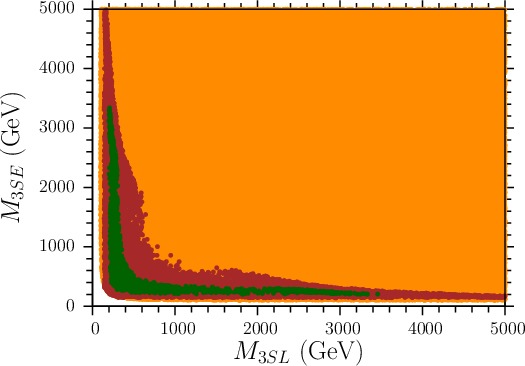}
\includegraphics[scale=0.4]{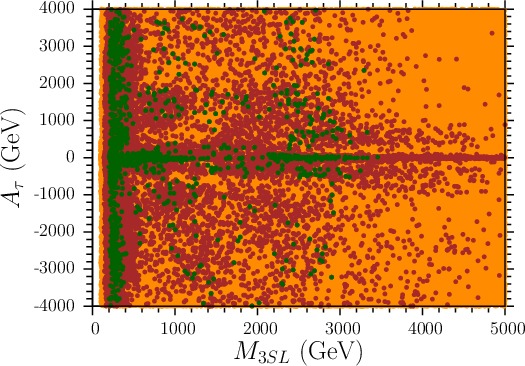}\vspace*{3mm}
\end{center}
\caption{$M_{3SE}$ vs. $M_{3SL}$ and $A_{\tau}$ vs. $M_{3SL}$ planes. Color coding is the same as in Figure \ref{fig:light-stop2}. }
\label{fig:only-light-stau2}
\end{figure}



\section{Light Stau in the Decoupling limit\label{sec:only-light-stau}}

One possible way of explaining the enhancement in the diphoton channel without significantly enhancing the gluon fusion rate and the decay widths of the other channels is by assuming the presence of a light stau \cite{Carena:2011aa}. In this section, we therefore assume the presence of light tau sleptons and study their effects Higgs production and decay in the decoupling limit. For this case we scan the parameter space as follows,
\begin{eqnarray}
   & 100\gev < M_{3SE}, M_{3SL} < 5000 \gev, \nonumber \\
   & {-4000\gev}< A_{\tau} < 4000\gev, \nonumber \\
   & 3 < \tan\beta  < 60.
\label{eq:param14}
\end{eqnarray}
As before, the first two generation masses are set equal to 5 TeV and all other A-terms are set to zero. The Higgs mass parameter $\mu$ and $m_A$ are also decoupled to 5 TeV.

 In Figure \ref{fig:only-light-stau1} we plot the same variables as in Figure \ref{fig:only-light-stop1} for the light stau scenario.  Figure \ref{fig:olta-a} shows an enhancement in \brhgm which increases for light stau masses. The branching ratio of the $ZZ$ channel is very close to its SM value (since $g \sin(\beta - \alpha) \sim 0$) as can be seen from  Figures \ref{fig:olta-b}. The stau with  no color charge does not affect the gluon fusion cross section. The effect on $Br(h \rightarrow ZZ)$ can be significant for low values of $m_A$ as we shall see in the next section. Earlier references (see for example, \cite{Carena:2011aa})  have noted that large values of the mixing parameter $A_{\tau}$ and moderate values of $m_A$ can lead to enhancement or suppression of the $h \rightarrow b \overline{b}$ decay which, in turn, can enhance or suppress the $h \rightarrow \gamma \gamma, WW $ and $ZZ$ decay modes. Moreover, the effects from a light stau can also be important in enhancing the diphoton branching ratio for suitably large values large $\mu \tan\beta$ \cite{Carena:2011aa}. Note that we do not apply the Higgs mass bound in this case because we set the mixing parameter $A_t=0$. Choosing suitably large values of this parameter or the  presence of vector like matter can accommodate the Higgs mass in the desired range given in Eq. (\ref{eq:mh-range}).

The plot of $R_{ ZZ} $ vs. $R_{ \gamma \gamma} $ in Figure \ref{fig:olta-f} shows that the product $R_{ ZZ} $ remains close to the SM value, whereas $R_{ \gamma \gamma} $ undergoes a strong enhancement. We can also derive an upper limit on the stau mass if the we require an enhancement in the diphoton channel as suggested by current observation. For $R_{\gamma \gamma} > 1.2$ the stau has to be $\lesssim 300 \gev$.
This scenario may be favored if in future analyses, the enhancement in the diphoton channel as seen by the CMS and ATLAS experiments persists, with the $ZZ$ channel being closer to its SM values. This can also be seen from the plot of the fundamental parameters in  Figure \ref{fig:only-light-stau2} where a large region of the parameter space satisfies the current limits from the ATLAS and CMS experiments. Moreover, the combination of the two experiments is also satisfied by a broad range of the parameter space.


\begin{figure}[]
\centering
\subfigure[]{\label{fig:lta-a}{\includegraphics[scale=0.4]{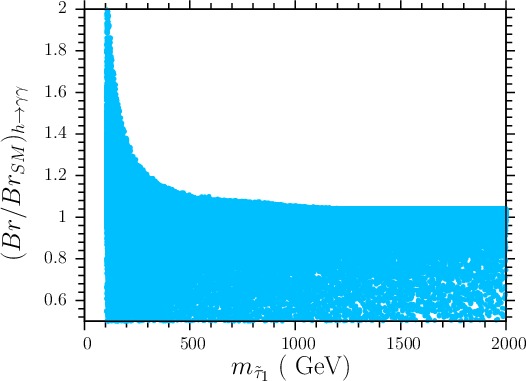}}}\hfill
\subfigure[]{\label{fig:lta-b}{\includegraphics[scale=0.4]{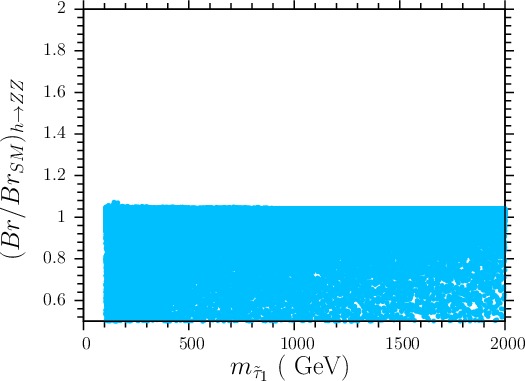}}}\\
\subfigure[]{\label{fig:lta-f}{\includegraphics[scale=0.41]{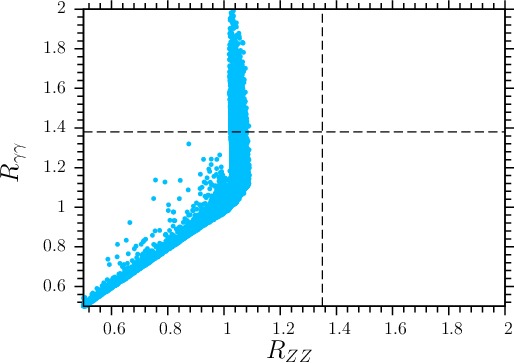}}}
\caption{$(Br/Br_{\rm SM})_{h \rightarrow \gamma \gamma}$ vs. $m_{\tilde{\tau}_1}$, $(Br/Br_{\rm SM})_{h \rightarrow ZZ}$ vs. $m_{\tilde{\tau}_1}$  and $(\sigma / \sigma_{\rm SM})_{gg \rightarrow h}$ vs. $m_{\tilde{t}_1}$ planes.
   Panel  \subref{fig:lta-f} shows the plot in the $R_{ \gamma \gamma}$ vs. $R_{ ZZ}$ planes.  The definition of the dashed lines is given in Figure \ref{fig:only-light-stop1}. All points satisfy the constraints described in section \ref{constraintsSection}.}
\label{fig:light-stau1}
\end{figure}



\begin{figure}[]
\centering
\subfigure[]{\label{fig:lta-2a}{\includegraphics[scale=0.4]{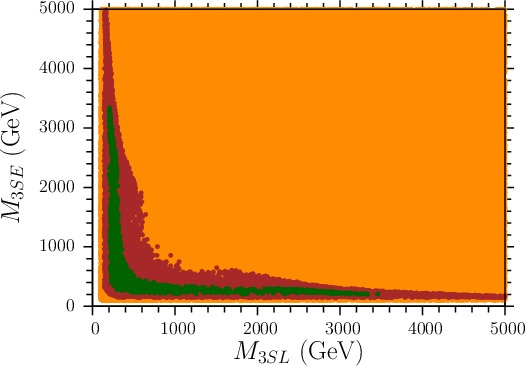}}}\hfill
\subfigure[]{\label{fig:lta-2b}{\includegraphics[scale=0.4]{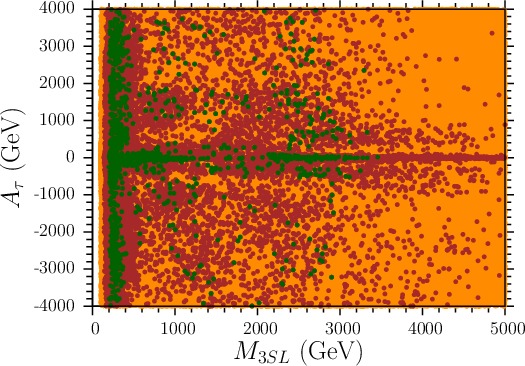}}}\\
\subfigure[]{\label{fig:lta-2c}{\includegraphics[scale=0.4]{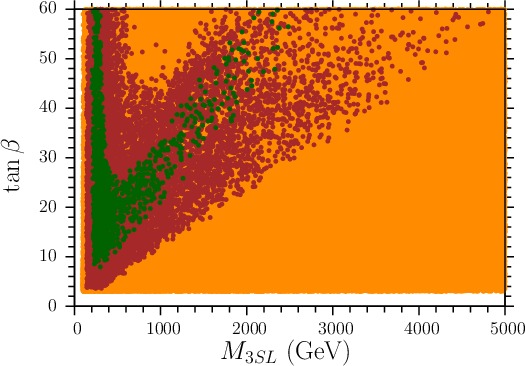}}}\hfill
\subfigure[]{\label{fig:lta-2d}{\includegraphics[scale=0.4]{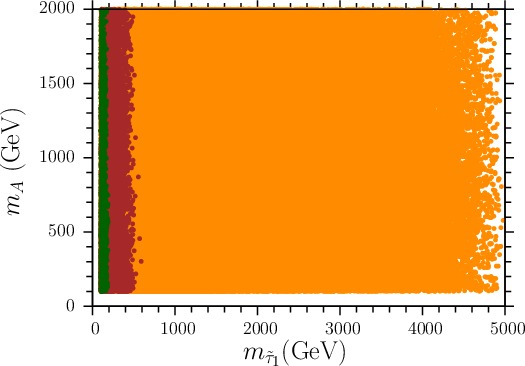}}}
\caption{$M_{3SE}$ vs. $M_{3SL}$, $A_{\tau}$ vs. $M_{3SL}$, $\tan\beta$ vs. $M_{3SL}$ and $m_A$ vs. $m_{\tilde{\tau}_1}$ planes. Color coding is the same as in Figure \ref{fig:light-stop2}.}
\label{fig:light-stau2}
\end{figure}



\section{Light Stau and low $m_A$ region\label{sec:light-stau}}

Our final scenario involves a light stau in the low $m_A$ region. For this case we scan the parameter space as follows:
\begin{eqnarray}
   & 100\gev < M_{3SE}, M_{3SL} < 5000 \gev, \nonumber \\
   & {-4000 \gev}< A_{\tau} < 4000\gev, \nonumber \\
 & 100\gev < m_A < 2000\gev, \nonumber \\
   & 100\gev <  \mu < 1000\gev, \nonumber \\
   & 3 < \tan\beta  < 60.
\label{eq:param15}
\end{eqnarray}
The first and second generation masses are decoupled to 5 TeV and all other A-terms are set to zero. The Higgs mass parameter $\mu$ and $m_A$ are also assumed to be 5 TeV. In Figure \ref{fig:lta-a} we can again see a large enhancement in the diphoton branching ratio for light stau masses. As described earlier, this enhancement corresponds to large values of $\mu \tan\beta$ as shown in previous references \cite{Carena:2011aa}. Our results are shown in Figures \ref{fig:light-stau1} and \ref{fig:light-stau2}. Comparing this case to the previous one we can see that the enhancement is not affected, whereas a large region of the parameter space now corresponds to suppressed values of the branching ratios for the $\gamma \gamma$ and $ZZ$ channels. As described in the previous section, lower/moderate values of $m_A$ can enhance $h \rightarrow b \overline{b}$ decay and therefore lead to a suppression of the diphoton and other decay channels. This suppression can be seen in Figures \ref{fig:lta-a}  and \ref{fig:lta-b}. The points with $Br \lesssim 1$ for the two channels correspond to $m_A \lesssim 700 \gev$.
The  cross section  is also suppressed for lower values of $m_A$.

 The plot of $R_{ZZ} $ vs. $R_{ \gamma \gamma} $ in Figure \ref{fig:lta-f} shows the enhancement seen in the previous case with additional suppression of the two channels corresponding to lower values of $m_A$. We can see that this case contributes more parameter space when  $R_{XX} \lesssim 1$  whereas the enhancement still corresponds to large values of $m_A$. For small values of the ratio $R$ there appears to be a linear relationship between the two products $R_{ZZ} $ and $R_{ \gamma \gamma} $ which is not present when $m_A$ is large. This scenario is therefore more favored compared to the previous one. 

  In Figure \ref{fig:light-stau2} plots in the fundamental parameter space plots further show the wide range of available parameter space  that satisfies the current constraints from experiments. The overlap of the green and brown points show that the combination of the two experiments is also broadly satisfied. Figure \ref{fig:lta-2d} shows that the current limits from the ATLAS and CMS experiments prefer a stau with mass $\lesssim 800 \gev$.


\section{Conclusion \label{conclusions}}

The ATLAS and CMS experiments have reported some exciting results regarding the production and subsequent decays (especially into $\gamma \gamma$ and $ZZ$) of a SM-like Higgs boson with mass close to 126 GeV. We have explored their implications in the MSSM framework with relatively light third generation sfermions (stop, sbottom and stau). We also considered scenarios in which TeV scale vector like particles are introduced to make sure that the Higgs boson has the desired mass of around 126 GeV. In addition, we explored both the decoupling limit ($m_A >> m_Z$) as well as the low $m_A$ region, with the first two family sfermions in all cases assumed to be essentially decoupled.

For the light stop case we find a wide region of the parameter space that can explain the current observations especially for low values of the pseudo-scalar mass $m_A$. Requiring the Higgs to be $126 \gev$ constrains the parameter space but the presence of vector-like matter can always ameliorate this situation. The case of light sbottom seems to be disfavored since the sbottom contribution in enhancement of the cross section and branching ratio is not large.
For the case of light staus we find a large region of the parameter space which agrees with current observations and also with the combined ATLAS and CMS limits.  More data from both experiments will help pin down the eventual scenario but, based on our analysis and also noted before by other authors, a light stau seems to be the most viable scenario in explaining current observed deviations from the SM.

\section*{Acknowledgments}
We thank   Sven Heinemeyer, Wolfgang Hollik  and Ian Low  for valuable discussions.
This work is supported in part by the DOE Grant No. DE-FG02-12ER41808. This work used the Extreme Science
and Engineering Discovery Environment (XSEDE), which is supported by the National Science
Foundation grant number OCI-1053575.

\newpage
\thispagestyle{empty}



\begin{thebibliography}{99}

\bibitem{ATLAS:jul4}
F. Gianotti, CERN Seminar, ``Update on the Standard Model Higgs searches in ATLAS",
July, 4 2012. ATLAS-CONF-2012-093

\bibitem{CMS:jul4}
J. Incandela, CERN Seminar, ``Update on the Standard Model Higgs searches in CMS",
July, 4 2012. CMS-PAS-HIG-12-020.

\bibitem{Tevatron:2012zz}
  The CDF and D0 Collaborations,
  FERMILAB-CONF-12-318-E.




\bibitem{Baer:2011ab} 
  H.~Baer, V.~Barger and A.~Mustafayev,
  Phys.\ Rev.\ D {\bf 85}, 075010 (2012);
   S.~Akula, B.~Altunkaynak, D.~Feldman, P.~Nath and G.~Peim,
  Phys.\ Rev.\ D {\bf 85}, 075001 (2012);
    A.~Arbey, M.~Battaglia, A.~Djouadi, F.~Mahmoudi and J.~Quevillon,
  Phys.\ Lett.\ B {\bf 708}, 162 (2012);
    J.~L.~Feng, K.~T.~Matchev and D.~Sanford,
  Phys.\ Rev.\ D {\bf 85}, 075007 (2012);
    S.~Heinemeyer, O.~Stal and G.~Weiglein,
  Phys.\ Lett.\ B {\bf 710}, 201 (2012);
      J.~Ellis   and K.~A.~Olive,
  Eur.\ Phys.\ J.\ C {\bf 72}, 2005 (2012);
    O.~Buchmueller, R.~Cavanaugh, A.~De Roeck, M.~J.~Dolan, J.~R.~Ellis, H.~Flacher, S.~Heinemeyer and G.~Isidori {\it et al.},
 arXiv:1112.3564 [hep-ph];
    M.~Kadastik, K.~Kannike, A.~Racioppi and M.~Raidal,
  JHEP {\bf 1205}, 061 (2012);
    J.~Cao, Z.~Heng, D.~Li and J.~M.~Yang,
  Phys.\ Lett.\ B {\bf 710}, 665 (2012);
     L.~J.~Hall, D.~Pinner and J.~T.~Ruderman,
  JHEP {\bf 1204}, 131 (2012);
     J.~Cao, Z.~Heng, J.~M.~Yang, Y.~Zhang and J.~Zhu,
  JHEP {\bf 1203}, 086 (2012);
  C.~-F.~Chang, K.~Cheung, Y.~-C.~Lin and T.~-C.~Yuan,
  JHEP {\bf 1206}, 128 (2012);
   L.~Aparicio, D.~G.~Cerdeno and L.~E.~Ibanez,
  JHEP {\bf 1204}, 126 (2012);
  H.~Baer, V.~Barger and A.~Mustafayev,
  JHEP {\bf 1205}, 091 (2012);
  D.~Ghosh, M.~Guchait and D.~Sengupta,
  arXiv:1202.4937 [hep-ph].
  L.~Maiani, A.~D.~Polosa and V.~Riquer,
  arXiv:1202.5998 [hep-ph];
  T.~Cheng, J.~Li, T.~Li, D.~V.~Nanopoulos and C.~Tong,
  arXiv:1202.6088 [hep-ph];
  J.~Cao, Z.~Heng, J.~M.~Yang and J.~Zhu,
  JHEP {\bf 1206}, 145 (2012);
  F.~Jegerlehner,
  arXiv:1203.0806 [hep-ph];
   F.~Brummer, S.~Kraml and S.~Kulkarni,
  arXiv:1204.5977 [hep-ph].
  A.~Choudhury and A.~Datta,
  JHEP {\bf 1206}, 006 (2012);
   C.~Balazs, A.~Buckley, D.~Carter, B.~Farmer and M.~White,
  arXiv:1205.1568 [hep-ph];
   M.~Badziak,
  Mod.\ Phys.\ Lett.\ A {\bf 27}, 1230020 (2012)
  [arXiv:1205.6232 [hep-ph]];
  J.~L.~Feng and D.~Sanford,
  arXiv:1205.2372 [hep-ph];
  E.~Dudas, Y.~Mambrini, A.~Mustafayev and K.~A.~Olive,
  arXiv:1205.5988 [hep-ph];
  A.~Fowlie, M.~Kazana, K.~Kowalska, S.~Munir, L.~Roszkowski, E.~M.~Sessolo, S.~Trojanowski and Y.~-L.~S.~Tsai,
  arXiv:1206.0264 [hep-ph];
   P.~Athron, S.~F.~King, D.~J.~Miller, S.~Moretti and R.~Nevzorov,
  arXiv:1206.5028 [hep-ph];
  R.~M.~Chatterjee, M.~Guchait and D.~Sengupta,
  arXiv:1206.5770 [hep-ph];
M.~W.~Cahill-Rowley, J.~L.~Hewett, A.~Ismail and T.~G.~Rizzo,
  arXiv:1206.5800 [hep-ph];
   S.~Akula, P.~Nath and G.~Peim,
  arXiv:1207.1839 [hep-ph];
  J.~Cao, Z.~Heng, J.~M.~Yang and J.~Zhu,
  arXiv:1207.3698 [hep-ph].

















\bibitem{Draper:2011aa}
  P.~Draper, P.~Meade, M.~Reece and D.~Shih,
  arXiv:1112.3068 [hep-ph];
Y.~Kats, P.~Meade, M.~Reece and D.~Shih,
  JHEP {\bf 1202}, 115 (2012);
  J.~L.~Evans, M.~Ibe, S.~Shirai and T.~T.~Yanagida,
  arXiv:1201.2611 [hep-ph].
  N.~Desai, B.~Mukhopadhyaya and S.~Niyogi,
  arXiv:1202.5190 [hep-ph];
  Z.~Kang, T.~Li, T.~Liu, C.~Tong and J.~M.~Yang,
  arXiv:1203.2336 [hep-ph].
 Z.~Kang, T.~Li, T.~Liu, C.~Tong and J.~M.~Yang,
  1203.2336 [hep-ph];
  M.A.~Ajaib, I.~Gogoladze, F.~Nasir and Q.~Shafi,
  1204.2856 [hep-ph];
  S.~P.~Martin and J.~D.~Wells,
  arXiv:1206.2956 [hep-ph];
   N.~Okada,
  arXiv:1205.5826 [hep-ph];
  J.~L.~Feng, Z.~'e.~Surujon and H.~-B.~Yu,
  arXiv:1205.6480 [hep-ph].
A.~Albaid and K.~S.~Babu,
  arXiv:1207.1014 [hep-ph].




\bibitem{Gogoladze:2011aa}
  I.~Gogoladze, Q.~Shafi and C.~S.~Un,
  arXiv:1112.2206 [hep-ph].
  I.~Gogoladze, Q.~Shafi and C.~S.~Un,
  JHEP {\bf 1207}, 055 (2012).

\bibitem{Chamseddine:1982jx}
 A.~Chamseddine, R.~Arnowitt and P.~Nath, Phys.\ Rev.\ Lett.\ {\bf 49} (1982) 970;
R.~Barbieri, S.~Ferrara and C.~Savoy, Phys.\ Lett.\ {\bf B119}
(1982) 343; N.~Ohta, Prog.\ Theor.\ Phys.\ {\bf 70} (1983) 542;
L.~J.~Hall, J.~D.~Lykken and S.~Weinberg, Phys.\ Rev.\ {\bf D27}
(1983) 2359; for a review see  S.~Weinberg, {\it The Quantum Theory
of Fields: Volume 3, Supersymmetry,
 Cambridge University Press (2000) 442p}.


\bibitem{big-422}
B. Ananthanarayan, G. Lazarides and Q. Shafi, Phys. Rev. D {\bf 44},
1613 (1991) and Phys. Lett. B {\bf 300}, 24 (1993)5; Q.~Shafi and
B.~Ananthanarayan, Trieste HEP Cosmol.1991:233-244.




\bibitem{CMS:ZZ-4l}
CMS Collaboration, CMS-PAS-HIG-12-016.




\bibitem{ATLAS:ZZ-4l}
ATLAS Collaboration, ATLAS-CONF-2012-092.




\bibitem{Arbey:2012dq}
  A.~Arbey, M.~Battaglia, A.~Djouadi and F.~Mahmoudi,
  arXiv:1207.1348 [hep-ph].







\bibitem{Djouadi:1998az}
  A.~Djouadi,
  Phys.\ Lett.\ B {\bf 435}, 101 (1998);
  R.~Dermisek and I.~Low,
  Phys.\ Rev.\ D {\bf 77}, 035012 (2008);
  I.~Low and S.~Shalgar,
  JHEP {\bf 0904}, 091 (2009).
  I.~Low, R.~Rattazzi and A.~Vichi,
  JHEP {\bf 1004}, 126 (2010);
 I.~Low and A.~Vichi,
  Phys.\ Rev.\ D {\bf 84}, 045019 (2011).



\bibitem{Desai:2012qy}
N.~Desai, B.~Mukhopadhyaya and S.~Niyogi,
  arXiv:1202.5190 [hep-ph];
 D.~Carmi, A.~Falkowski, E.~Kuflik and T.~Volansky,
  arXiv:1202.3144 [hep-ph];
  N.~D.~Christensen, T.~Han and S.~Su,
  arXiv:1203.3207 [hep-ph];
  R.~Benbrik, M.~G.~Bock, S.~Heinemeyer, O.~Stal, G.~Weiglein and L.~Zeune,
  arXiv:1207.1096 [hep-ph];
  N.~Arkani-Hamed, K.~Blum, R.~T.~D'Agnolo and J.~Fan,
  arXiv:1207.4482 [hep-ph];
  A.~Joglekar, P.~Schwaller and C.~E.~M.~Wagner,
  arXiv:1207.4235 [hep-ph];
 M.~R.~Buckley and D.~Hooper,
  arXiv:1207.1445 [hep-ph];

  
\bibitem{Carena:2011aa} 
  M.~Carena, S.~Gori, N.~R.~Shah and C.~E.~M.~Wagner,
  JHEP {\bf 1203}, 014 (2012)
  M.~Carena, S.~Gori, N.~R.~Shah, C.~E.~M.~Wagner and L.~-T.~Wang,
  arXiv:1205.5842 [hep-ph];
  G.~F.~Giudice, P.~Paradisi and A.~Strumia,
  arXiv:1207.6393 [hep-ph].



  %
\bibitem{Moroi:1991mg}
  T.~Moroi, Y.~Okada,
  Mod.\ Phys.\ Lett.\  {\bf A7 } (1992)  187;
  Phys.\ Lett.\  {\bf B295 } (1992)  73.

\bibitem{Gogoladze:2009bd}
K.~S.~Babu, I.~Gogoladze, C.~Kolda,
  [hep-ph/0410085]
  K.~S.~Babu, I.~Gogoladze, M.~U.~Rehman and Q.~Shafi,
  Phys.\ Rev.\ D {\bf 78}, 055017 (2008);
  S.P.~Martin,
  Phys.\ Rev.\ D {\bf 81}, 035004 (2010);Phys.\ Rev.\ D {\bf 82}, 055019 (2010)
   P.~W.~Graham, A.~Ismail, S.~Rajendran and P.~Saraswat,
  Phys.\ Rev.\ D {\bf 81}, 055016 (2010);
T.~Li, J.~A.~Maxin, D.~V.~Nanopoulos and J.~W.~Walker,
  Phys.\ Lett.\ B {\bf 710}, 207 (2012);
M.~Endo, K.~Hamaguchi, S.~Iwamoto and N.~Yokozaki,
  Phys.\ Rev.\ D {\bf 85}, 095012 (2012).



\bibitem{Anastasiou:2002yz}
  C.~Anastasiou and K.~Melnikov,
  Nucl.\ Phys.\ B {\bf 646}, 220 (2002).


\bibitem{Djouadi:2005gj}
  A.~Djouadi,
  Phys.\ Rept.\  {\bf 459}, 1 (2008).



\bibitem{gunion90}
J.~F.~Gunion, H.~E.~Haber, G.~L.~Kane and S.~Dawson,
{\it``The Higgs Hunter's Guide''}, Addison-Wesley, Reading (USA), 1990.
%

\bibitem{Boonekamp:2005up} 
  M.~Boonekamp, J.~Cammin, S.~Lavignac, R.~B.~Peschanski and C.~Royon,
  Phys.\ Rev.\ D {\bf 73}, 115011 (2006)
  [hep-ph/0506275].

\bibitem{feynhiggs}
  M.~Frank, T.~Hahn, S.~Heinemeyer, W.~Hollik, H.~Rzehak and G.~Weiglein,
  JHEP {\bf 0702}, 047 (2007);
%
  G.~Degrassi, S.~Heinemeyer, W.~Hollik, P.~Slavich and G.~Weiglein,
  Eur.\ Phys.\ J.\ C {\bf 28}, 133 (2003);
%
  S.~Heinemeyer, W.~Hollik and G.~Weiglein,
  Eur.\ Phys.\ J.\ C {\bf 9}, 343 (1999);
%
  S.~Heinemeyer, W.~Hollik and G.~Weiglein,
  Comput.\ Phys.\ Commun.\  {\bf 124}, 76 (2000).












\bibitem{:1900yx}
 [Tevatron Electroweak Working Group and CDF Collaboration and D0 Collab],
  arXiv:0903.2503 [hep-ex].


\bibitem{Leva}
J.L. Leva, A fast normal random number generator, ACM Trans. Math. Softw. 18 (1992) 449-453;
J.L. Leva, Algorithm 712. A normal random number generator, ACM Trans. Math. Softw. 18 (1992) 454-455.


\bibitem{Aad:2011ib}
  G.~Aad {\it et al.} [ ATLAS Collaboration ],
  arXiv:1109.6572 [hep-ex].


\bibitem{He:2011tp}
M.~A.~Ajaib, T.~Li and Q.~Shafi,
  Phys.\ Rev.\ D {\bf 85}, 055021 (2012);
  B.~He, T.~Li and Q.~Shafi,
  JHEP {\bf 1205}, 148 (2012).





\bibitem{arXiv:1103.4344}
  G.~Aad {\it et al.} [ATLAS Collaboration],
  arXiv:1103.4344 [hep-ex].



\bibitem{AdeelAjaib:2011ec}
  M.~Adeel Ajaib, T.~Li and Q.~Shafi,
  Phys.\ Lett.\ B {\bf 701}, 255 (2011).




\bibitem{Amsler:2008zzb}
  K. Nakamura {\it et al.} [ Particle Data Group Collaboration ],
  J.\ Phys.\ G {\bf G37}, 075021 (2010).




\bibitem{Jungman:1995df}
  G.~Jungman, M.~Kamionkowski and K.~Griest,
  Phys.\ Rept.\  {\bf 267}, 195 (1996).



\bibitem{at}
Y.~Okada, M.~Yamaguchi and T.~Yanagida, Prog.\ Theor.\ Phys.\  {\bf
85}, 1 (1991);
Phys.\ Lett.\ B {\bf 262}, 54 (1991);
  A.~Yamada,
  Phys.\ Lett.\  B {\bf 263}, 233 (1991);
J.R.~Ellis, G.~Ridolfi and F.~Zwirner, Phys.\ Lett.\ B {\bf 257}, 83
(1991);
Phys.\ Lett.\ B {\bf 262}, 477 (1991);
H.E.~Haber and R.~Hempfling, Phys.\ Rev.\ Lett.\ {\bf 66},  1815
(1991).


\bibitem{Carena:1995wu}
M.~Carena, J.~R.~Espinosa, M.~Quiros and C.~E.~M.~Wagner,
Phys.\ Lett.\ B {\bf 355}, 209 (1995);
M.~Carena, M.~Quiros and C.~E.~M.~Wagner,
Nucl.\ Phys.\ B {\bf 461}, 407 (1996);
H.~E.~Haber, R.~Hempfling and A.~H.~Hoang,
Z.\ Phys.\ C {\bf 75}, 539 (1997);
S.~Heinemeyer, W.~Hollik, and G.~Weiglein,
Phys.\ Rev.\ D {\bf 58}, 091701 (1998);
M.~Carena, H.~E.~Haber, S.~Heinemeyer, W.~Hollik, C.~E.~M.~Wagner,
and G.~Weiglein,
Nucl.\ Phys.\ B {\bf 580}, 29 (2000);
S.~P.~Martin,
Phys.\ Rev.\ D {\bf 67}, 095012 (2003).

\bibitem{Carena:1998gk}
  M.~S.~Carena, S.~Mrenna and C.~E.~M.~Wagner,
  Phys.\ Rev.\ D {\bf 60}, 075010 (1999).

\bibitem{Carena:1999bh}
  M.~S.~Carena, S.~Mrenna and C.~E.~M.~Wagner,
  Phys.\ Rev.\ D {\bf 62}, 055008 (2000);
  J.~Cao, Z.~Heng, T.~Liu and J.~M.~Yang,
  Phys.\ Lett.\ B {\bf 703}, 462 (2011).
  





\end{thebibliography}
\end{document}